\documentclass[sigconf, nonacm]{acmart}

\usepackage{algorithm}
\usepackage{algorithmic}
\usepackage[table]{xcolor}
\usepackage{tikz}
\usepackage{pgfplots}
\usepackage{multirow}
\usepackage{subcaption}

\usetikzlibrary{trees,positioning,arrows,decorations.pathreplacing,shapes,calc}
\usepackage{graphicx}
\usepackage{booktabs}

\microtypesetup{protrusion=true,expansion=true}
\begin{document}
\title[TabTracer: MCTS for Complex Table Reasoning with LLMs]{TabTracer: Monte Carlo Tree Search for Complex Table Reasoning with Large Language Models}


\author{Zhizhao Luo$^{1,2}$, Zhaojing Luo$^2$, Meihui Zhang$^{2, *}$, Rui Mao$^3$}
\affiliation{
  \institution{$^1$Beijing Institute of Technology, Zhuhai, $^2$Beijing Institute of Technology, $^3$Shenzhen University}
  \country{} 
}
\email{{zzluo, zjluo, meihui_zhang}@bit.edu.cn, mao@szu.edu.cn}

\renewcommand{\shortauthors}{Luo et al.}

\begin{abstract}

Large language models (LLMs) have emerged as powerful tools for natural language table reasoning, where there are two main categories of methods.
Prompt-based approaches rely on language-only inference or one-pass program generation without step-level verification.
Agent-based approaches use tools in a closed loop, but verification is often local, and backtracking is limited, allowing errors to propagate and increasing cost.
Moreover, they rely on chain- or beam-style trajectories that are typically combinatorially redundant, leading to high token costs.
In this paper, we propose TabTracer, an agentic framework that coordinates multi-step tool calls over intermediate table states, with explicit state tracking for verification and rollback. 
First, it enforces step-level verification with typed operations and lightweight numeric and format checks to provide reliable rewards and suppress hallucinations. 
Second, execution-feedback Monte Carlo Tree Search maintains a search tree of candidate table states and uses backpropagated reflection scores to guide UCB1 selection and rollback via versioned snapshots. 
Third, it reduces redundancy with budget-aware pruning, deduplication, and state hashing with a monotonicity gate to cut token cost. 
Comprehensive evaluation on TabFact, WikiTQ, and CRT datasets shows that TabTracer outperforms state-of-the-art baselines by up to 6.7\% in accuracy while reducing token consumption by 59--84\%. 

\end{abstract}

\maketitle



\section{Introduction}
Structured tables are a ubiquitous medium for organizing data~\cite{borisovlanguage,gorishniy2021revisiting} across domains such as healthcare analytics~\cite{cai2024cohortnet}, financial risk management~\cite{altman2023realistic}, and scientific discovery~\cite{burdick2020table}. 
However, their explicit structure often imposes rigid constraints that are misaligned with the flexibility of natural language, making it difficult to map ambiguous user intents to precise table operations for compositional inference~\cite{fanglarge}. 
To address these complexities, table reasoning studies how to answer natural-language questions grounded in tabular evidence~\cite{pasupat2015compositional}.

Early table reasoning methods primarily relied on executable semantic parsing or program induction~\cite{pasupat2015compositional,pasupat2016inferring}. 
Although interpretable, they were burdened by heavy symbolic engineering and search, and were fragile under weak supervision, non-canonical cell values, and schema/domain shifts~\cite{jauhar2016tables,zhang2017macro}. 
Table-oriented pretraining~\cite{yin2020tabert,herzig2020tapas} was introduced to ease this engineering burden and robustness fragility by learning stronger text-table representations, enabling end-to-end reasoning without explicit logical forms. 
Yet this line still depended on task-specific fine-tuning, so generalization across tasks and domains remained constrained. 
By contrast, large language models (LLMs)~\cite{yang2025qwen3,hurst2024gpt} offer stronger in-context generalization and broader cross-schema transfer.

\begin{figure}[tb]
    \centering
    \includegraphics[width=\linewidth]{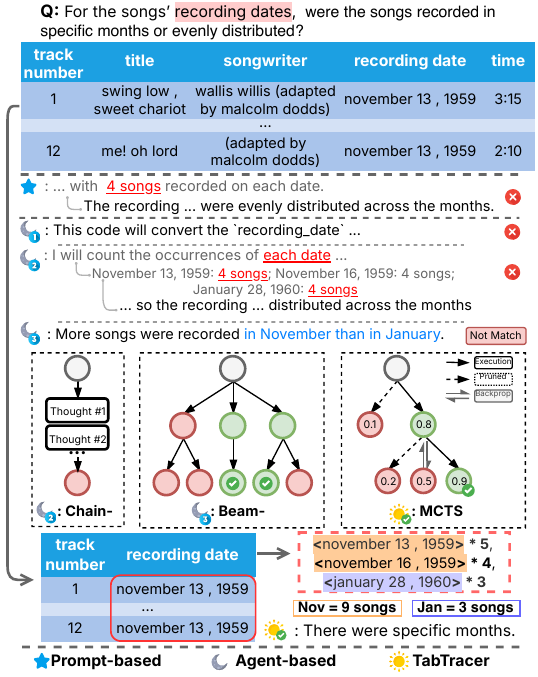}
    \vspace{-4ex}
    \caption{Prompt-based and agent-based outputs fail to complete the aggregation, while TabTracer(our approach) slices the table to count songs per date and aggregate by month (Nov=9 vs Jan=3).}
    \Description{Two table segments labelled Case A and Case B, each comparing prompt-based, agent-based, and TabTracer conclusions with icons for NH, FH, and PED.}
    \vspace{-4ex}
    \label{fig:intro}
\end{figure}

LLM-based table reasoning treats the LLM as the core reasoning engine~\cite{shankar2024building} and relies on prompting to drive table understanding, reasoning, and answering. 
Early LLM-based approaches are largely prompt-based~\cite{chen2023large,wang2023plan,zhouleast}. They feed the table and question in a single shot or a fixed multi-stage prompt, where the step order is pre-defined, and stage outputs are passed forward in one direction, without changing downstream decisions based on intermediate execution. 
There is no external execution signal to confirm whether intermediate hypotheses (e.g., chosen columns or filtered rows) are correct, so once a model misreads the schema or keeps the wrong subset, subsequent reasoning is built on that error. In Figure~\ref{fig:intro}, CoT~\cite{wei2022chain} (star) illustrates how language-only judgment can introduce counting or statistical errors that then propagate along the chain. 
These limitations motivate agent-based methods~\cite{guo2024large,yaoreact,zhang2024chain}, which close the loop by letting the LLM (or multiple LLMs) select actions, call tools, and update decisions based on execution feedback. 
However, the feedback is often local, systematic backtracking is limited, and multi-branch reasoning inflates token cost. 
In Figure~\ref{fig:intro}, ReAcTable~\cite{reactable2024} (moon-1) fails to trigger tool use and stalls before code generation, Chain-of-Table~\cite{chainoftable2024} (moon-2) accumulates process errors without timely rollback, misreading month as ``date'' and failing to aggregate by month, and Table-Critic~\cite{tablecritic2025} (moon-3) is close but mismatched in expression while incurring heavy token cost from multi-branch dialogue. 
Overall, existing methods still face clear challenges in both accuracy and efficiency.

\textbf{Challenge 1:} 
Existing methods lack step-level verification and reliable rewards, so intermediate mistakes often go unchecked and propagate. 
When tables are linearized for LLMs, two-dimensional structural cues such as row and column alignment and schema, type, or unit constraints are weakened~\cite{li2024table,papotti2025panel}, which makes semantic errors and tabular hallucinations harder to detect. 
LLMs are also insensitive to numerical precision in counting and aggregation tasks, leading to numerical hallucinations~\cite{chen2025expanding,ke2024unveiling} that further compound across steps. 
We need machine-checkable step verification and execution-grounded rewards that reflect intent, where lightweight numeric and format checks~\cite{caciularu2024tact} make verification practical and keep reasoning grounded in evidence~\cite{psallidas2018smoke,reactable2024,chainoftable2024}. 
This is difficult because table operators and data formats vary widely, making it hard to define unified pre- and post-conditions~\cite {shinn2023reflexion,liuagentbench}. 
It is also hard to build reliable rewards when many tool interfaces are free-form or weakly parsed, which makes intermediate verification unreliable.
 
\textbf{Challenge 2:} 
Lack of effective backtracking makes early errors hard to correct. 
Most methods follow a linear or shallow branching process, so once the question or table is misread early, later steps can only patch the current path locally~\cite{guo2026rethinking}. 
They rarely roll back key decisions or explore alternatives, which entrenches and amplifies errors. 
A better approach is a backtrackable reasoning mechanism that can return to key decision points when execution evidence conflicts with the current hypothesis, reset or replace subpaths, and choose among candidate branches based on evidence~\cite{puttaswamy2025delta}. 
However, this is challenging. Execution feedback is often local and noisy, which makes it hard to form stable value estimates or confidence scores for reliable backtracking, and the system must balance continuing the current path against rolling back to explore alternatives to avoid oscillation or over-backtracking.

\textbf{Challenge 3:} 
There is no budget-aware search that suppresses redundant branches while keeping progress stable within token and latency constraints. Many multi-branch or multi-agent methods adopt chain-style or beam-style search topologies~\cite{wan2024alphazero,yao2023tree,chen2024tree}. Chain-style reasoning, as shown in Figure~\ref{fig:intro}, cannot backtrack, so an early mistake collapses the entire path. Beam search in Figure~\ref{fig:intro} retains multiple branches, but many are near duplicates, leading to redundant expansions and high token and execution costs~\cite{zhuang2024toolchain}. A more robust strategy involves execution-grounded prioritization, pruning, and deduplication, which collectively ensure search efficacy within a constrained budget~\cite{haffner2023efficiently,chen2023loger}.
However, implementing this approach is fraught with practical difficulties. One primary obstacle stems from the inherent noise and locality of execution feedback. Such limitations can lead pruning heuristics to inadvertently discard valid reasoning paths while retaining misleading ones.
Furthermore, preventing the search from revisiting redundant table states presents a significant challenge. For instance, alternating operations of table shrinkage and expansion can cause the search process to degenerate into cycles~\cite{zhou2024language}. While structuring the search space as a Directed Acyclic Graph (DAG) is theoretically ideal for efficiency, ensuring such acyclicity remains difficult to guarantee in practical implementations.

In this paper, we propose TabTracer, which is an agentic framework that runs an execute-and-reflect loop over intermediate table states. 
The system starts with lightweight profiling of the table and question, extracting column names and types, missing-value patterns, table scale, and question type to seed the initial state and constrain the action space. 
First, to enforce step-level verification and provide execution-grounded rewards, TabTracer employs typed table operators for column selection, row filtering, and computation, and each operation produces a versioned intermediate table accompanied by machine-checkable pre- and post-checks. This mechanism ensures that the reasoning process remains grounded in verifiable evidence, thereby effectively suppressing numerical hallucinations and preventing error propagation across steps.
Secondly, to enable reliable backtracking, execution-feedback Monte Carlo Tree Search(MCTS)~\cite{browne2012mcts} maintains a tree of candidate table states, backpropagates reflection scores to update node values for UCB1 selection, and relies on versioned snapshots to support rollback and subpath replacement when evidence conflicts. 
Thirdly, to control redundancy and token cost, we enforce a fixed token budget with pruning and deduplication. State hashing and reuse suppress near-duplicate expansions, and a monotonicity gate commits a node only when the semantic state genuinely changes, which yields DAG-like progress and reduces redundant executions under the budget. 
As shown in Figure~\ref{fig:intro}, TabTracer first slices a focused subtable (e.g., track/recording date) and then performs targeted aggregations via state-validated tool executions to yield the final answer. The resulting search tree is more compact, reaches the answer with fewer expansions, and implies lower token usage.

In summary, our main contributions are as follows:
\begin{itemize}
\item We present TabTracer, an agentic framework that uses typed operators with step-level checks to produce reliable execution rewards and reduce hallucinations.
\item We develop an execution-feedback MCTS framework that maintains candidate table states and supports evidence-driven backtracking with versioned snapshots, correcting early errors rather than patching the current path. 
\item We propose budget-aware pruning and deduplication with state hashing, monotonicity gating, and failure memory to suppress near-duplicate expansions and reduce token consumption under a fixed budget. 
\item We demonstrate through comprehensive evaluation on TabFact, WikiTQ, and CRT datasets that TabTracer outperforms state-of-the-art baselines by up to 6.7\% in accuracy while reducing token consumption by 59--84\%.
\end{itemize}

The remainder of the paper is organised as follows: Section 2 reviews related work, Section 3 elaborates on the TabTracer methodology, Section 4 presents experimental settings and results, and Section 5 concludes the paper.
\section{Related Work}
\label{sec:related}
\vspace{-1ex}
\subsection{Prompt-Based Reasoning Methods}
\label{subsec:prompt-based}
Prompt-based methods organize the table and question into a single-shot or fixed two-stage prompt. The LLM does not make multi-step decisions based on external execution feedback, so the reasoning relies on language-only inference or one-pass program generation. 
We group this line into language-only self-reasoning and fixed-pipeline programs.

\textbf{Language-only self-reasoning. }
These methods avoid executable programs and external feedback, and instead use prompt structure and representation to guide reasoning. Tab-CoT~\cite{ziqi2023tab} aligns rows and columns with a structured Markdown reasoning table. Two-stage self-augmentation prompting extracts structural cues before reasoning and improves robustness~\cite{sui2024table}. Simple table serialization yields better table-reasoning performance with minimal in-context examples~\cite{chen2023large}. FLEXTAF~\cite{zhang2024flextaf} improves alignment via multi-format selection and voting. This line mainly improves prompt structure, representation, and stability, but still lacks tool-level verification and rollback, so early deviations are hard to correct. In this work, TabTracer grounds language reasoning in executable tools and adds step-level verifiable feedback and reflection-based rewards, reducing the accumulation of format-correct but semantically wrong steps.

\textbf{Fixed-pipeline programs. }
These methods generate a program or extraction directive once, and the execution result does not feed back into subsequent decisions, forming a one-way pipeline. Binder~\cite{chengbinding} embeds LLM APIs inside hybrid programs, but the execution flow remains fixed. TabSQLify~\cite{nahid2024tabsqlify} operates on a SQL-to-subtable-to-answer workflow, whereas H-STAR~\cite{abhyankar2025h} utilizes a two-stage approach comprising fixed extraction and routed reasoning. LOTUS~\cite{lotus2024} introduces semantic operators with query optimization. TabTrim~\cite{guo2026rethinking} performs gold-trajectory supervised pruning with parallel subtable configurations. These methods heavily depend on the semantic correctness of extraction or program generation, and once a subtable or program is wrong, recovery is difficult. In contrast, TabTracer replaces the one-way pipeline with information-gain MCTS and uses versioned table states with hash-based deduplication to enable backtracking and reuse.



\vspace{-1ex}
\subsection{Agent-Based Reasoning Systems}
\label{subsec:agent-based}

Agent-based methods treat the LLM as a controller that makes multi-step decisions in a closed loop, using tools, retrieval, or execution feedback to update subsequent actions. We group them into tool-using agents and multi-agent collaboration.

\textbf{Tool-using agents.}
This line focuses on repeated tool calls with feedback to improve executability and control. Chain-of-Table~\cite{chainoftable2024} evolves the table via iterative operations. ReAcTable~\cite{reactable2024} brings ReAct to table QA and uses execution feedback with voting across multiple trajectories. AixelAsk~\cite{zhang2025aixelask} formulates reasoning as DAG optimization with decomposition and retrieval. TableMind~\cite{jiang2025tablemind} trains tool-use policies with SFT and RAPO. Despite feedback, verification often stays at syntax or local execution, with limited semantic checks or backtracking. In this work, TabTracer adds step-level machine-checkable verification and reflection rewards, and couples tool execution with MCTS-style backtracking to correct early mistakes.

\textbf{Multi-agent collaboration. }
These methods emphasize coordination among multiple agents that share context, critique, or iteratively refine results. AutoTQA~\cite{zhu2024autotqa} uses multi-role FSM scheduling for multi-table QA. MACT~\cite{mact2025} introduces online planning. Table-Critic~\cite{tablecritic2025} relies on multi-role critique with a self-evolving template tree. TALON~\cite{jin2025talon} adopts an exploration-retrieval-answer pipeline for long tables. 
However, verification is still largely language-based, multi-agent dialogue inflates token cost and latency, and systems often repair within the current path rather than systematically backtrack. Our approach avoids multi-agent overhead by using budget-aware search with state reuse, and suppresses redundant branches via a blacklist of known bad state-action pairs to maintain stable exploration at lower cost.

\begin{figure*}[t]
  \centering
  \includegraphics[width=\textwidth]{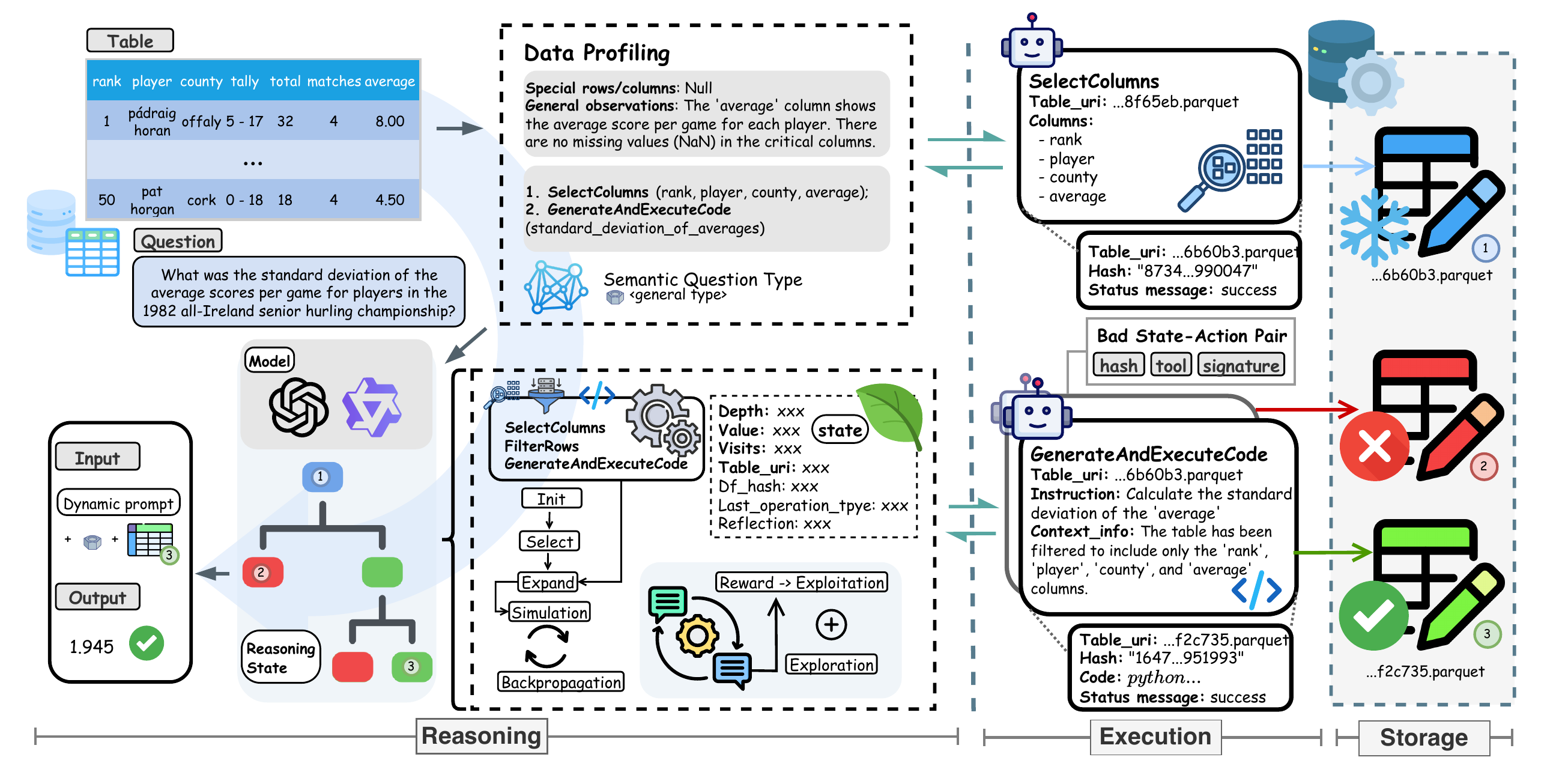}
  \vspace{-6ex}
  \caption{The reasoning layer includes planning and reflection, the execution layer issues atomic dataframe tools, and the versioned storage layer preserves snapshots for fallback and retry. }
  \label{fig:architecture}
  \vspace{-4ex}
\end{figure*}
\vspace{-1ex}
\section{Method}
\label{sec:method}

To address three challenges in table reasoning, namely weak step-level verification and unreliable rewards that lead to hallucinations, lack of effective backtracking, and redundant search cost, TabTracer is an agentic framework that couples LLM-based planning with environment interaction over semantic table states, while explicitly managing those states for verification and recovery.
As shown in Figure~\ref{fig:architecture}, it first uses the \textbf{Reasoning Layer} to run budget-aware MCTS that manages branching, enables backtracking, and suppresses near-duplicate paths. 
The selected actions are then executed by the \textbf{Execution Layer} via typed dataframe tools with pre/post-condition checks. Finally, the \textbf{Storage Layer} preserves versioned table states so rollback and reuse remain possible.

Formally, we denote the layered agent as:
\begin{equation}
\setlength{\abovedisplayskip}{4pt}
\setlength{\belowdisplayskip}{4pt}
\mathcal{G} = (\mathcal{L}_{\mathrm{reason}}, \mathcal{L}_{\mathrm{exec}}, \mathcal{L}_{\mathrm{store}}),
\end{equation}
where $\mathcal{L}_{\mathrm{reason}}$ hosts the planner, $\mathcal{L}_{\mathrm{exec}}$ exposes deterministic dataframe tools (column selection, row filtering, and code-based computation), and $\mathcal{L}_{\mathrm{store}}$ provides a hash-indexed rollback cache that stores versioned table snapshots for reuse and rollback. 
The subsequent subsections elaborate on each layer before analysing their joint guarantees and runtime optimisations. 


\subsection{Reasoning Layer}
\label{subsec:reasoning}
This layer begins with table-aware data profiling before running the budgeted MCTS controller and reflection loop that decide which tool call to issue next. 
Accordingly, we first describe the profiling routine and its induced state abstraction, then detail the information-guided MCTS update in Section~\ref{subsec:mcts}, and finally show how the resulting scores underpin convergence guarantees relied upon by the subsequent layers.
\subsubsection{\textbf{Data profiling.}}
The left block of Figure~\ref{fig:architecture} shows the data profiling. We load the full table into a dataframe, log row/column counts, detect special rows/columns or missing values (e.g., whether the \textit{average} column is complete), and render a markdown preview that becomes the planner’s contextual memory. In parallel, the question is rephrased and classified (numerical, comparative, lookup, etc.), and the profiler proposes an initial tool skeleton. 
This metadata materialises the root snapshot—capturing the table URI, semantic hash, planner/tool traces, and the latest reflection—and is sufficient for downstream planning.

Here a \emph{snapshot} denotes the persisted table state after a tool execution (identified by its URI and hash), while a \emph{trace} records the reasoning messages that produced that state. 
Formally, we view TabTracer as an MDP $\mathcal{M}=\langle \mathcal{S},\mathcal{A},\mathcal{T},\mathcal{R},\pi\rangle$ whose state space collects those persisted snapshots and their semantic traces:
\begin{equation}
  \mathcal{S} = \big\{ \langle u, \chi, \mu, r \rangle : u \in \mathcal{U},\; \chi \in \mathcal{H},\; \mu \in \mathcal{M}_{\text{msg}},\; r \in \mathcal{R}_{\text{refl}} \big\}.
\end{equation}
Here $u$ is the URI of a Parquet snapshot, $\chi$ is its semantic hash (a deterministic fingerprint derived from the dataframe content) returned by deterministic tools, $\mu$ stores the ordered planner/tool messages that produced the state (serving as the semantic context previously denoted by $\mathcal{C}$), and $r$ is the most recent reflection summarising execution feedback. The finite action set is $\mathcal{A}=\{\textsc{FilterRows},\textsc{SelectColumns},\textsc{GenExeCode}\}$. 
Here \textsc{SelectColumns} reduces column width, \textsc{FilterRows} prunes rows with guarded predicates, and \textsc{GenExeCode} runs aggregations or multi-step computations. Details are given in Section~\ref{subsec:execution}. Actions correspond to deterministic tools. 
Transition $\mathcal{T}$ loads the snapshot denoted by $u_t$, applies the requested tool, and writes the successor back to storage together with its hash. Reward $\mathcal{R}$ is the normalised reflection score. This compact record is Markov-sufficient because deterministic tool execution implies:
\begin{equation}
  P(s_{t+1}, r_t \mid s_{0:t}, a_{0:t}) = P(s_{t+1}, r_t \mid s_t, a_t),
  \label{eq:markov}
\end{equation}
so the semantic hash acts as a certificate that two nodes encode the same dataframe, while the message/reflection traces preserve the semantic context needed for subsequent prompts. Rather than maintaining a full hash ledger, TabTracer stores only the current snapshot plus a cache of ineffective state--action signatures, which keeps the planner observable without bloating memory.
This state abstraction supports downstream planning and makes the stored snapshots sufficient for subsequent tool use.

\subsubsection{\textbf{Information-Guided MCTS Cycle.}}
\label{subsec:mcts}
We define the search tree over table states. Each node stores a snapshot together with its reasoning trace, and each edge corresponds to a single action that transforms the table state, typically via a tool invocation.
Sequential execution without backtracking creates brittle reasoning chains where early errors corrupt downstream computations. TabTracer employs information-guided MCTS to maintain multiple candidate paths and enable recovery from suboptimal decisions. 
Guided by the state view above, the planner repeats four stages per simulation: (i) \textbf{Selection} walks the existing tree with an exploration-biased score.
(ii) \textbf{Expansion} executes the most promising normalised tool proposal under deterministic guards. (iii) \textbf{Reflection} queries the reflection operator for dense feedback about the newly created node, and (iv) \textbf{Backup} propagates the scalar reward and solved status to the root. The complete procedure is formalized in Algorithm~\ref{alg:tabtracer}.
To exploit the structured state, the planner executes a single tool per simulation, obtains dense rewards through reflection, and prioritises nodes via the score in Eq.~\eqref{eq:ucb}. 

Although backpropagation uses the scalar reward $r_t$, the rollout-level objective aggregates multiple execution signals. Let $\phi_t = (\phi^{\text{row}}_t, \phi^{\text{col}}_t, \phi^{\text{rel}}_t, \phi^{\text{sol}}_t)$ collect row/column shrinkage, tool relevance, and solution-confidence heuristics derived from execution metadata. The rollout objective is then:
\begin{equation}
  \mathcal{J}(\pi) = \mathbb{E}_{\tau \sim \pi}\left[ \sum_{t=0}^{T-1} \gamma^t \,f_{\text{eval}}(\phi_t) \right],
  \label{eq:trajectory-objective}
\end{equation}
where $f_{\text{eval}}$ denotes the reflection scorer that integrates these signals. The reflection module receives all components of $\phi_t$ via structured prompts (including row/column counts from Eq.~\eqref{eq:rho}, tool metadata, and question context), enabling context-dependent integration. The resulting reflection score $r_t$ represents an implicit evaluation $r_t \approx f_{\text{eval}}(\phi_t)$, where the LLM adaptively weights features based on question type and data characteristics.

For fallback ranking, we also maintain penalty indicators for duplicate or degenerate states:
\begin{equation}
  \Phi_{\text{pen}}(\tau) = \sum_{t=0}^{T-1} \gamma^t \Big( \lambda_{\text{dup}} \mathbf{1}_{\text{dup}}(s_{t+1}) + \lambda_{\text{empty}} \mathbf{1}_{\text{empty}}(s_{t+1}) \Big),
  \label{eq:exec-penalty}
\end{equation}
where $\mathbf{1}_{\text{dup}}$ detects hash-duplicate states ($\delta=0$ in Eq.~\eqref{eq:delta}) and $\mathbf{1}_{\text{empty}}$ flags degenerate tables. These penalties are evaluated lazily and used by the fallback scorer described later.


In effect, TabTracer keeps the planning loop analytically lightweight, while embedding execution-aware heuristics exactly where they matter, inside the rollout objective, so the scheduler remains theoretically clean and practically data-efficient.


\begin{algorithm}[tb]
  \caption{TabTracer Single-Step MCTS}
  \label{alg:tabtracer}
  \begin{algorithmic}[1]
    \REQUIRE Initial state $s_0$, simulation budget $B$, blacklist $\mathcal{K}$
    \STATE Initialise root node $n_0$ with $s_0$
    \FOR{$t = 1$ to $B$}
    \STATE $n \leftarrow$ \textsc{SelectLeaf}$(n_0)$ \COMMENT{UCB1 selection with guards}
    \STATE $(child, a, s', \rho_{n,a}) \leftarrow$ \textsc{ExpandNode}$(n, \mathcal{K})$ \COMMENT{Alg.~\ref{alg:expand}}
    \IF{$child = \bot$}
      \STATE \textbf{continue}
    \ENDIF
    \STATE $(score, solved, evidence) \leftarrow$ \textsc{Reflect}$(n.state, a, s')$ \COMMENT{Alg.~\ref{alg:reflect}}
    \STATE \textsc{Backpropagate}$(child, score, solved, \rho_{n,a}, evidence)$ \COMMENT{update value/visits and solved flag}
      \IF{$solved$}
        \STATE \textbf{break}
      \ENDIF
    \ENDFOR
    \RETURN \textsc{BestSolution}$(n_0)$
  \end{algorithmic}
\end{algorithm}
Algorithm~\ref{alg:tabtracer} summarizes the single-step MCTS loop. Lines 2--3 initialize the root state and iterate over the simulation budget. Each iteration selects a leaf (line 4), expands it with one tool action (line 5), reflects on the outcome (line 6), and backpropagates the score (line 7). If a solution is found, the loop terminates early (lines 8--9), the search continues until the budget is exhausted and returns the best solution (line 11). 

\textbf{Selection stage.}
In Figure~\ref{fig:architecture}, selection picks the leaf with the highest UCB1 score among current candidates. For this query, that leaf typically retains the \textit{average} column, so subsequent actions can compute the standard deviation. 
The selection procedure uses a lightweight UCB1 score that retains the simplicity of UCB1 while incorporating execution-aware guards:
\begin{equation}
  \text{UCB1}(c)=c.value + C\sqrt{\frac{\log (n.visits+1)}{c.visits+\epsilon}},
  \label{eq:ucb}
\end{equation}

where $c$ is a child of $n$ (the parent), $C$ is the exploration constant, and $\epsilon$ avoids division by zero. Children with zero visits receive effectively infinite priority, so the search quickly touches unexplored branches before refining high-value ones. Selection iteratively descends from the root by choosing the child with the highest UCB1 score among eligible nodes and stops at the first unvisited node. After this scoring step, TabTracer applies execution-aware guards—hash-based duplicate detection, empty-table rollback, and monotonicity gates—to ensure only genuinely informative states remain on the frontier.


\textbf{Expansion stage.}
For each selected leaf, 
For each selected leaf node, which represents the current table snapshot, the expansion procedure (Algorithm~\ref{alg:expand}) uses the LLM to propose an ordered list of candidate tool actions, applies deterministic validity filters (column existence, argument validity, cached bad state--action signatures, duplicate/ineffective actions, and tool guards), and executes the first valid action to create a child node. We define the semantic state change indicator.
In Figure~\ref{fig:architecture}, the planner proposes multiple candidate actions, then filters invalid or low-quality ones before executing the first valid tool call, such as \textsc{SelectColumns} to keep \{rank, player, county, average\}. After execution, the resulting subtable is saved as a new snapshot, which becomes the child node for the next stage.
\begin{equation}
  \delta(D,D') = \begin{cases} 1 & \text{if } \chi(D') \neq \chi(D) \\ 0 & \text{otherwise} \end{cases}
  \label{eq:delta}
\end{equation}
where $\chi$ denotes the content hash. The shrinkage proxy then tracks data reduction:
\begin{equation}
  \rho(D,a) = \delta(D,D') \cdot \phi_{\text{rows}}(D,D') \cdot \phi_{\text{cols}}(D,D'),
  \label{eq:rho}
\end{equation}
where $\phi_{\text{rows}}(D,D') = \text{rows}(D')/\text{rows}(D)$ and $\phi_{\text{cols}}(D,D') = \text{cols}(D')/\text{cols}(D)$ denote relative dimension changes. The implementation uses $\delta$ as a hard monotonicity gate: child nodes are created only when $\delta=1$, ensuring that every expansion alters the table state. Dimension ratios are tracked as metadata and provided to the reflection module for quality assessment.

Because the executor only materialises child nodes when the semantic hash changes, every accepted expansion corresponds to a semantically updated table state. This monotonicity helps the planner avoid redundant revisits. If an action leaves the table unchanged ($\delta=0$, hence $\rho=0$), it is immediately discarded. We still impose fixed depth and simulation budgets as hard safeguards, but the shrinkage signal gives us a direct, execution-time measurement for ranking expansions by how much evidence they retain.

Algorithm~\ref{alg:expand} details this step. It first asks the LLM to propose a small ordered set of candidate actions for the current state (line 2). Each candidate is normalized into a concrete tool call (line 4), filtered by validity checks and cached bad signatures (lines 5--8), and the first candidate that passes these checks is executed to produce a new snapshot and child node (lines 9--11). Candidates that fail validation are skipped.
\begin{algorithm}[t]
  \caption{\textsc{ExpandNode}}
  \label{alg:expand}
  \begin{algorithmic}[1]
    \REQUIRE node $n$, blacklist $\mathcal{K}$, candidate budget $k$
    \STATE $\mathcal{A}_{\text{cand}} \leftarrow \textsc{LLMPropose}(n.state, k)$ \COMMENT{ordered candidates}
    \FOR{$a$ in $\mathcal{A}_{\text{cand}}$}
      \STATE $(tool, args) \leftarrow \textsc{NormalizeAndCoerce}(a)$
      \IF{$tool = \varnothing$} \STATE \textbf{continue} \ENDIF
      \IF{\textsc{ColumnOrArgInvalid}(tool, args, n)} \STATE \textbf{continue} \ENDIF
      \IF{$(n.hash, tool, args) \in \mathcal{K}$} \STATE \textbf{continue} \ENDIF
      \IF{\textsc{DuplicateOrIneffective}(n, tool, args)} \STATE \textbf{continue} \ENDIF
      \IF{\textsc{ViolatesToolGuards}(tool, args, n)} \STATE \textbf{continue} \ENDIF
      \STATE $s' \leftarrow \mathcal{T}(n.state, tool, args)$
      \STATE $child \leftarrow \textsc{NewNode}(s', \text{metadata})$
      \STATE Append $child$ to $n.children$
      \RETURN $(child, tool, s', \text{metadata})$
    \ENDFOR
    \RETURN $(\bot, \varnothing, \varnothing, 1)$
  \end{algorithmic}
\end{algorithm}

Candidate selection follows ordered proposals with explicit validity gates:
\begin{equation}
\setlength{\abovedisplayskip}{5pt}
\setlength{\belowdisplayskip}{5pt}
  \mathcal{A}_{\text{cand}} = [a_1,\ldots,a_k] \sim p_{\theta}(\cdot \mid s_t),
  \label{eq:action-ranking}
\end{equation}
\begin{equation}
  \mathcal{A}_{\text{valid}}(s_t)=\{a\in\mathcal{A}_{\text{cand}}:\textsc{Valid}(a,s_t)\},
  \label{eq:valid-gate}
\end{equation}
\begin{equation}
  i^*=\min\{i\in\{1,\ldots,k\}:a_i\in\mathcal{A}_{\text{valid}}(s_t)\},\quad a_t=a_{i^*}.
  \label{eq:first-valid}
\end{equation}
where the LLM proposal order provides the implicit priority, and deterministic filters enforce executability. This realizes a constrained first-valid selection rule consistent with the implementation.
Here $p_{\theta}(\cdot\mid s_t)$ denotes the LLM-induced proposal distribution conditioned on state $s_t$, and \textsc{Valid}$(a,s_t)$ denotes deterministic executability checks including column existence, argument validity, bad-pair filtering, duplicate/ineffective filtering, and tool guards.

\textbf{Reflection stage.}
For the hurling query, reflection verifies that the new snapshot still contains the \textit{average} column and is suitable for computing its standard deviation, then assigns a score. Candidates that drop required columns or yield unusable tables are recorded as bad state--action pairs. 
The reflective evaluator consumes $(s,a,s')$ plus row/column diffs and previews, returning only a normalised scalar score $r \in [0,1]$, a boolean solved flag, and the free-form critique stored in each node. This dense reward is immediately backpropagated, so the online MCTS loop maintains exactly the statistics described in Algorithm~\ref{alg:reflect}. Execution metadata (hashes, row counts, tool names) is cached alongside the reflection so that the fallback scorer can later recover richer signals without perturbing the online propagation. When the tree needs to surface a final answer, we therefore assemble a lazily evaluated feature vector from the cached metadata and reflections:
When ranking frontier nodes, the system constructs a feature vector:
\begin{equation}
  \label{eq:reward-vector}
  \boldsymbol{z}_t = [\,r_t,\; \delta_t,\; \phi^{\text{row}}_t,\; \phi^{\text{col}}_t,\; \mathbf{1}_{\text{solve}}\,]^{\top},
\end{equation}
comprising normalized reflection score $r_t$, state change indicator $\delta_t$, dimension changes, and solution status. These components are provided to the reflection module via structured prompts (including tool outputs and metadata), enabling integrated quality assessment. The accumulated node value $V_n$, obtained through backpropagation (Eq.~\ref{eq:backprop}), represents this integrated score. Fallback selection then uses:
\begin{equation}
  n^* = \arg\max_{n \in \mathcal{N}_{\text{frontier}}} V_n - \beta_{\text{pen}} \cdot \mathbf{1}_{\text{degenerate}}(n),
  \label{eq:fallback-selection}
\end{equation}
where $\mathcal{N}_{\text{frontier}}$ denotes the current frontier node set, and the penalty term suppresses empty or duplicate states.

Algorithm~\ref{alg:reflect} makes this process explicit: it builds a structured prompt from $(s,a,s')$, calls the reflection operator, and returns the score, solved flag, and critique used for backpropagation. 
\begin{algorithm}[t]
  \caption{\textsc{Reflect}}
  \label{alg:reflect}
  \begin{algorithmic}[1]
    \REQUIRE state $s$, action $a$, successor $s'$
    \STATE $prompt \leftarrow$ \textsc{ComposePrompt}$(s,a,s')$
    \STATE $resp \leftarrow$ \textsc{ReflectOperator}$(prompt)$ \COMMENT{Structured output: reflections, score, flag}
    \IF{$resp = \varnothing$}
      \STATE \textbf{return} $(0, \text{False}, \text{``Reflection unavailable.''})$
    \ENDIF
    \STATE $score \leftarrow \mathrm{clip}(resp.score/10, 0, 1)$
    \STATE $solved \leftarrow resp.found\_solution$
    \RETURN $(score, solved, resp.reflections)$
  \end{algorithmic}
\end{algorithm}

Reflection, therefore, produces only a normalised score, a boolean flag, and the critique attached to each node. Degeneracy checks (hash-duplicate pruning, empty-table rollback, tool guards) run before reflection, and richer heuristics such as shrinkage or solution-confidence cues are read off the cached metadata during fallback. 

\textbf{Backup stage.}
Backup propagates the reflection score and solved flag upward to update node values and visit counts for UCB1 selection and enable early termination. 
The backpropagated score updates the node value (an incremental mean of reflection scores) and visit count along the root path, which raises the UCB1 score for branches whose snapshots retain the evidence needed for the query (e.g., the \textit{average} column). This makes later selections more likely to revisit those branches, where the next expansion can invoke \textsc{GenExeCode} to compute the final statistic and start a new iteration of the cycle.
The backpropagation procedure propagates rewards upwards using incremental means:
Starting from the expanded node, we update its visit count and value using the incremental mean in Eq.~\eqref{eq:backprop}, then repeat this update along the parent chain until reaching the root. If the reflection marks the node as solved, we propagate the solved flag upward, which can trigger early termination.
\begin{equation}
\setlength{\abovedisplayskip}{5pt}
\setlength{\belowdisplayskip}{5pt}
  V_n \leftarrow V_n + \frac{R_{\text{child}} - V_n}{N_n},
  \label{eq:backprop}
\end{equation}
where $R_{\text{child}}$ is the normalized reflection reward returned by the expanded child and $N_n$ is the updated visit count of node $n$.



Each node also stores lightweight execution metadata (last tool type, output rows, hashes) that the fallback scorer later consumes when ranking leaves: the enhanced heuristic combines these cached counters with the final reflections to recover shrinkage, relevance, and solution-confidence terms even though the online backpropagation keeps only $(value, visits)$. The search depth is bounded by the explicit depth cap and the shrinkage rate,
\begin{equation}
  t \;\le\; \min\!\left(d_{\max},\ \log_{1/\eta}\bigl(\min(R_0,C_0)\bigr)\right),
\end{equation}
and admissible trajectories are those that pass step-level tool guards,
\begin{equation}
  \mathbf{1}_{\text{adm}}(\tau)=\prod_{t=0}^{T-1}\mathbf{1}\bigl[\textsc{Guards}(s_t,a_t)=\text{True}\bigr],
\end{equation}
so invalid actions are filtered immediately, while hash gating prevents revisiting identical table states.

\subsubsection{\textbf{Answer Selection and Termination.}}
The reasoning loop terminates either when the root is marked solved (a child has found a certificate) or when the simulation budget is exhausted. In both cases, Eq.~\eqref{eq:reward-vector} provides a lazily evaluated feature vector that ranks frontier nodes by accumulated reflection scores, shrinkage, and degeneracy penalties while suppressing state--action signatures already blacklisted. The heavy lifting—storing hashes, metadata, and penalty weights—is handled by the storage layer (Section~\ref{subsec:storage}); the reasoning layer simply queries this cache to obtain $(\boldsymbol{z}_t, s^{\text{fb}}_t)$ and selects the highest-ranked node for answer extraction. 
\subsection{Execution Layer}
\label{subsec:execution}
This layer corresponds to the middle block in Figure~\ref{fig:architecture}: every action chosen by the planner is grounded in one of three typed dataframe operators whose outputs are machine-checkable tuples $(\texttt{table\_uri}, \texttt{hash}, \texttt{rows}, \texttt{cols}, \texttt{preview}, \texttt{status})$. The goal is to turn abstract instructions into deterministic table manipulations while streaming back the metadata that drives shrinkage signals and reflection prompts.
Our tool design focuses on progressively shrinking the table while retaining key evidence, so we use column selection and row filtering as the primary reduction steps. The remaining fine-grained format handling and numerical computation are consolidated into code generation, which keeps the tool set minimal while still supporting complex calculations with verifiable, traceable outputs.
\subsubsection{\textbf{Deterministic dataframe operators}}
The execution layer grounds reasoning in concrete table transformations rather than language-only guesses. A compact motivation is that deterministic transformations do not increase information about the answer,
\begin{equation}
  I(A;D' \mid Q) \le I(A;D \mid Q),
\end{equation}
and retaining both snapshots preserves information,
\begin{equation}
  H(A \mid D, D', Q) = H(A \mid D, Q).
\end{equation}
Here, $A$ is the answer, $Q$ the question, $D$ the current table, and $D'$ the table after a deterministic transformation.
The three tools in Figure~\ref{fig:architecture} implement these maps: \textsc{SelectColumns} projects the schema to a minimal column set, \textsc{FilterRows} evaluates guarded predicates, and \textsc{GenExeCode} runs sandboxed Python for bespoke statistics. Each tool emits the tuple mentioned above so the planner sees the updated snapshot and preview. 

\subsubsection{\textbf{Typed interfaces and validation}}
LLM proposals are first normalised by a tool-call normalizer that coerces JSON payloads and fills in missing IDs. Each deterministic tool then applies schema-aware guards—checking column existence, predicate validity, type/format coercion, and sandbox constraints—before execution, and emits explicit status codes and diagnostics. Rich execution metadata (row/column counts, hashes, previews) returns with every call so the reasoning layer can compute shrinkage, detect degeneracy, and compose reflection prompts, while the same tuple feeds the storage layer for versioning. The \textit{bad state–action pair} panel in Figure~\ref{fig:architecture} visualises how signatures are logged whenever validation fails. 
For compactness, Algorithm~\ref{alg:tool-interfaces} summarises these contracts end-to-end.
SelectColumns performs case-insensitive column matching and returns a versioned snapshot with row/column counts and a preview. FilterRows normalises predicates and types before applying the query, then emits the filtered snapshot and metadata. GenExeCode generates and executes Python for aggregations or multi-step arithmetic when selection and filtering are insufficient, returning the result snapshot with execution metadata.
\begin{algorithm}[t]
  \caption{Typed tool contracts. Each call consumes a validated snapshot and emits a versioned successor.}
  \label{alg:tool-interfaces}
  \begin{algorithmic}[1]
    \STATE \textbf{SelectColumns}($table\_uri, columns[\,]$)
    \STATE \quad $D \leftarrow \textsc{LoadDF}(table\_uri)$
    \STATE \quad $columns_{\text{actual}} \leftarrow \textsc{GuardColumns}(D, columns)$
    \STATE \quad $D' \leftarrow D[columns_{\text{actual}}]$
    \STATE \quad \textbf{return} $(uri', hash', rows(D'), cols(D'), preview, status)$
    \STATE
    \STATE \textbf{FilterRows}($table\_uri, condition$)
    \STATE \quad $D \leftarrow \textsc{LoadDF}(table\_uri)$
    \STATE \quad $D_{\text{safe}} \leftarrow \textsc{TypeNormalise}(D)$
    \STATE \quad $D' \leftarrow \textsc{EvaluatePredicate}(D_{\text{safe}}, condition)$
    \STATE \quad \textbf{return} $(uri', hash', rows(D'), cols(D'), preview, status)$
    \STATE
    \STATE \textbf{GenExeCode}($table\_uri, instruction, context$)
    \STATE \quad $D \leftarrow \textsc{LoadDF}(table\_uri)$
    \STATE \quad $code \leftarrow \textsc{LLMGenerate}(instruction, context, D)$
    \STATE \quad $D' \leftarrow \textsc{SandboxExecute}(code, D)$
    \STATE \quad $\textsc{QualityGuards}(D')$
    \STATE \quad \textbf{return} $(uri', hash', rows(D'), cols(D'), preview, code, status)$
  \end{algorithmic}
\end{algorithm}

\subsubsection{\textbf{Execute-and-reflect convergence}}
Because each simulation executes exactly one deterministic tool call followed by the reflection operator, the per-iteration cost is dominated by the execution-layer operations described in Section~\ref{subsec:execution}. Under the standard UCB assumptions (bounded rewards and persistent exploration) and consistent reflection scores, TabTracer's selection rule satisfies
\begin{equation}
\setlength{\abovedisplayskip}{5pt}
\setlength{\belowdisplayskip}{5pt}
  \lim_{t \to \infty} P(\pi_t = \pi^*) = 1,\label{eq:convergence}
\end{equation}
\begin{equation*}
\setlength{\abovedisplayskip}{5pt}
\setlength{\belowdisplayskip}{5pt}
  \mathbb{E}\left[ V^{\pi^*}(s_0) - V^{\pi_t}(s_0) \right] = O\!\left(\sqrt{\frac{\log t}{t}}\right).
\end{equation*}
The classic UCB1 confidence radius directly applies here. Immediate reflection yields unbiased value estimates conditioned on observed transitions (due to Eq.~\eqref{eq:markov}), and single-step expansions avoid simulation bias. Plugging these elements into the standard UCB1 regret decomposition establishes the bound.
This result indicates that TabTracer's single-step execute-and-reflect loop approaches high-quality policies while keeping runtime predictable, meaning the execution layer (deterministic tool + reflection) preserves the exploration--exploitation trade-off implied by Eq.~\eqref{eq:convergence}.

\subsection{Storage Layer}
\label{subsec:storage}
This layer matches the rightmost block in Figure~\ref{fig:architecture}: every tuple returned by the execution layer is materialised as a versioned snapshot and hash-indexed cache entry, with penalties recorded for ineffective state–action pairs. 
Its role is twofold—reuse past evidence via cached snapshots and suppress redundant or low-quality expansions via deduplication and blacklists. 
We next describe evidence caching and reuse, followed by memory/penalty management.

\subsubsection{\textbf{Evidence cache and reuse}}
Each snapshot is materialised as the tuple $v_i = (\texttt{uri}_i, \texttt{hash}_i, \mu_i)$, where $\mu_i$ tracks row/column counts and execution metadata. Rather than storing full trajectories, TabTracer maintains a hash-indexed cache: every semantic hash points to its Parquet URI, the accompanying metadata, and two signature-based filters. First, an \emph{action deduplication} set suppresses semantically equivalent tool calls from the same state (avoid re-expanding duplicate actions). Second, a \emph{blacklist} stores state--action signatures that previously yielded low reflection scores or invalid outputs; these pairs are penalised and skipped in later rollouts. Penalty scores assigned to $(h,a)$ pairs that repeatedly failed act as soft constraints in subsequent rollouts, while the cached URI lets fallback routines dereference $v_i$ without re-running the tool chain.

Let $\mathcal{H}_t$ denote the hash-indexed cache up to time $t$, where $h$ denotes a semantic table-state hash. Whenever TabTracer executes $(h,a)$ and the reflection score remains below tolerance, the pair is recorded with a penalty weight $\pi(h,a)$:
\begin{equation}
  \mathcal{H}_{t+1} = \mathcal{H}_t \cup \bigl\{ (h,a,\pi(h,a)) : \pi(h,a) < 0 \bigr\}.
\end{equation}
Revisiting the same $h$ first queries $\mathcal{H}_t$: signatures with accumulated penalties are skipped, whereas unexplored signatures can reuse the stored snapshot. The hash-aware pruning, therefore, prevents repeated low-quality expansions and concentrates budget on novel, higher-utility actions.

In addition to pruning, the cache exposes previously solved URIs for fallback routines. Figure~\ref{fig:architecture} depicts this behaviour: red crosses mark signatures that have accumulated penalties (hence future rollouts skip them), whereas green check marks indicate snapshots that can be dereferenced without re-running tools. Because each entry retains the Parquet pointer and table metadata, rereading a hash only requires dereferencing $v_i$ rather than replaying the entire tool chain. The storage layer thus remains lightweight: it stores snapshots, penalties, and re-usable URIs, but no full reward histories or posterior parameters.

\subsubsection{\textbf{Memory Management}}
Only well-typed actions (indicator $\mathbf{1}_{\text{typed}}(a,D)=1$) are dispatched, and any failure signature $(h_t,a)$ is appended to the blacklist $\mathcal{K}$ so subsequent attempts are dropped without re-running tools (here $h_t$ is the current state's semantic hash). The storage layer combines lazy loading and hash-based caching—if two nodes share the same semantic hash, only pointers are retained. Incremental statistics such as the mean value in Eq.~\eqref{eq:backprop} avoid storing full reward histories. Low-utility branches are pruned when their UCB1 score drops below
\begin{equation}
  \xi_t = \max_{n \in \text{frontier}} \text{UCB1}(n,t) - \kappa,
  \label{eq:prune}
\end{equation}
with $\kappa$ calibrated to keep the tree within a fixed memory budget. 

\section{Experiments}
\label{sec:experiments}
In this section, we present the experimental evaluation of TabTracer. We first introduce the datasets, baselines, and evaluation metrics used in our experiments. We then report results on overall effectiveness and cost-efficiency against strong baselines. We additionally analyze numerical robustness on arithmetic-heavy workloads, conduct tool-level ablations, and further examine scalability with respect to table size and search depth.

\subsection{Experimental Setup}
\label{subsec:setup}
\subsubsection{Datasets.}
We evaluate TabTracer on the official test sets of TabFact, WikiTQ, and CRT. 
Table~\ref{tab:datasets} shows the statistical results of the datasets we used. 
TabFact contains 2,024 table-claim pairs, each pair answered with a binary label (Yes/No). 
WikiTQ covers 4,344 open-domain questions whose answers can be strings, numbers, or dates. 
CRT comprises 728 open-ended tasks spanning factual and analytical queries, where 204 of them (28.0\%) require explicit numerical computation. 

\begin{table}[tb]
  \centering
  \caption{Datasets used in the experiments. ``Numeric'' denotes the percentage of questions whose gold answers contain numeric values. Table (S/M/L) corresponds to table size <500, 500--1500, and >1500 tokens.}
  \label{tab:datasets}
  \vspace{-2ex}
  \small
  \setlength{\tabcolsep}{0.2mm}
    \begin{tabular}{lcccc}
      \toprule
      Dataset & \# Queries & Numeric & Answer Type & Table (S/M/L) \\
      \midrule
      TabFact & 2,024 & -- & Yes / No & 1188 / 813 / 23 \\
      WikiTQ  & 4,344 & 51.3\% & String / Number / Date & 2006 / 1867 / 471 \\
      CRT     & 728   & 28.0\% & Free-form (text + numeric) & 499 / 221 / 8 \\
      \bottomrule
    \end{tabular}
    \vspace{-3ex}
\end{table}

\subsubsection{System Configuration}
\label{subsec:system}
\paragraph{Experimental Environment.} Experiments run on a Linux~6.6.80 host with a 16-core Intel Xeon Gold~6133 CPU and 31~GiB RAM. 
Qwen2.5-72B-Instruct~\cite{qwen2.5} is served via vLLM~\cite{kwon2023efficient} on 8 NVIDIA V100 GPUs (CUDA~12.1). 
Qwen3-32B~\cite{qwen3technicalreport} leverages 8 NVIDIA L40 GPUs (CUDA~12.1). 
Qwen3-32B-Direct disables the provider-side \textit{thinking} pattern, thus acting as a pure instruction-following model without auxiliary reasoning tokens. 
GPT-4.1-mini~\cite{openai2025gpt41} is accessed through the OpenAI API. 
The implementation relies on PyTorch~2.1~\cite{paszke2017automatic}, langgraph~0.5.0, langchain~0.3.26, and pandas~2.3.0. 

\paragraph{TabTracer Implementation.} We generate three candidate expansions per node, cap the search depth at $5$, set the simulation budget to $15$, and apply a LangGraph recursion limit of $50$. 
Hash-based state caching and rollback are enabled by default. 
The base temperature is set to $0.1$ for candidate generation and reflection. 

\paragraph{Metrics.} 
We report Exact Match (EM), token input/output counts, tool Success Rate (SR), tool adoption rate (AR, i.e., the fraction of tool invocations whose outputs remain on retained branch), state reuse rate, and average simulation count. 
We follow the Table-Critic protocol to normalise punctuation, formats, and timestamps before comparison. 
For WikiTQ and CRT, we compare answers after normalization, whereas TabFact directly uses EM.

\subsubsection{Baselines}
\label{subsubsec:baselines}

We chose a comprehensive set of nine baseline models for comparison. 
All baselines are evaluated under the same experimental conditions (model version, hardware, and evaluation metrics) to ensure a fair comparison. 
Hyperparameters are tuned on validation sets with systematic grid search to optimize performance while maintaining comparable computational budgets.

\paragraph{Prompt-Based Reasoning Methods:}
\begin{itemize}
\item \textbf{End-to-End QA}: Direct question-answering without intermediate steps, using standard few-shot prompting with 0 exemplar per dataset.
\item \textbf{Few-Shot QA}: Enhanced few-shot prompting with 3-5 carefully selected exemplars covering different question types and table structures.
\item \textbf{Chain-of-Thought (CoT)} \cite{wei2022chain}: Standard CoT prompting with step-by-step reasoning for table reasoning tasks.
\item \textbf{CoT-Consistent} \cite{cotc2024}: Self-consistency CoT with $k=5$ samples and majority voting for final answer selection.
\item \textbf{Lotus} \cite{lotus2024}: Semantic operator framework with declarative DataFrame operations and statistical accuracy guarantees. We implement the core semantic operators (sem\_filter, sem\_join, sem\_topk) and integrate domain-specific constraints for financial and scientific schemas.
\end{itemize}

\paragraph{Agent-Based Reasoning Systems:}
\begin{itemize}
\item \textbf{Chain-of-Table} \cite{chainoftable2024}: Deterministic transformation programs with operation classifiers retrained on our datasets. Maximum program length capped at 12 operations to prevent infinite loops.
\item \textbf{Reactable}\cite{reactable2024}: ReAct-style agent with iterative reasoning-action loops. Configuration includes full tool suite (filter, sort, aggregate, join) with beam search (beam size = 4) for action selection.
\item \textbf{MACT} \cite{mact2025}: Multi-agent collaborative framework with planner, executor, and verifier agents. We employ the recommended two-round deliberation protocol and share our tool adapters across all agents for consistency.
\item \textbf{Table-Critic} \cite{tablecritic2025}: Iterative self-critique system with three critique-refinement loops. It uses the same operator set as Chain-of-Table for tool execution.
\end{itemize}

\vspace{-2.5ex}
\subsection{Comparison with Baselines}
\label{subsec:overall}

\begin{table*}[tb]
\centering
\caption{Experimental Results on Table QA Datasets, where results show Exact Match accuracy. Bold numbers denote the best results. Underline numbers imply the second-best results.}
\vspace{-2ex}
\label{tab:main_experiment}
\resizebox{\textwidth}{!}{%
\begin{tabular}{l|ccc|ccc|ccc|ccc}
\toprule
\multirow{2}{*}{\textbf{Method}} & \multicolumn{3}{c|}{\textbf{Qwen3-32B-Direct}} & \multicolumn{3}{c|}{\textbf{Qwen2.5-72B}} & \multicolumn{3}{c|}{\textbf{Qwen3-32B}} & \multicolumn{3}{c}{\textbf{GPT-4.1-Mini}} \\
 & TabFact & WikiTQ & CRT & TabFact & WikiTQ & CRT & TabFact & WikiTQ & CRT & TabFact & WikiTQ & CRT \\
\midrule
\multicolumn{13}{c}{\textit{Prompt-Based Reasoning}} \\
\midrule
End-to-End QA & 0.7915 & 0.5615 & 0.4904 & 0.8557 & 0.6271 & 0.5275 & 0.9407 & 0.7585 & 0.6093 & 0.8597 & 0.6634 & 0.5508 \\
Few-Shot QA & 0.8187 & 0.5930 & 0.5137 & 0.8439 & 0.5907 & 0.4835 & 0.9303 & 0.7742 & 0.6058 & 0.8295 & 0.5654 & 0.4629 \\
Chain-of-Thought & 0.9185 & 0.5413 & 0.4698 & 0.9175 & 0.7671 & 0.6181 & 0.9298 & 0.7546 & 0.6202 & 0.9437 & 0.7581 & 0.5865 \\
CoT-Consist & 0.9338 & 0.5927 & 0.4959 & \underline{0.9358} & \underline{0.7828} & \underline{0.6305} & \underline{0.9447} & 0.7789 & 0.6168 & \textbf{0.9491} & 0.7726 & 0.6016 \\
Lotus & 0.8434 & 0.5750 & 0.4533 & 0.8869 & 0.6722 & 0.5659 & 0.9224 & 0.7516 & 0.6140 & 0.8908 & 0.7319 & 0.5659 \\
\midrule
\multicolumn{13}{c}{\textit{Agent-Based Systems}} \\
\midrule
Chain-of-Table & 0.8864 & 0.4918 & 0.4317 & 0.9077 & 0.7231 & 0.5604 & 0.9030 & 0.7540 & 0.4684 & 0.8664 & 0.6237 & 0.5483 \\
Reactable & 0.8063 & 0.6001 & 0.5618 & 0.6764 & 0.5320 & 0.3571 & 0.8167 & 0.5992 & 0.5591 & 0.6809 & 0.4936 & 0.2816 \\
MACT & 0.8706 & 0.7120 & 0.5824 & 0.8923 & 0.7306 & 0.5934 & 0.9422 & 0.7957 & \underline{0.6621} & 0.8923 & 0.6685 & 0.5797 \\
Table-Critic & \underline{0.9430} & \underline{0.8301} & \underline{0.6414} & 0.9260 & 0.7770 & 0.6250 & 0.9417 & \underline{0.8285} & 0.6428 & 0.9402 & \underline{0.8128} & \underline{0.6634} \\
 \textbf{TabTracer (Ours)} & \textbf{0.9447} & \textbf{0.8503} & \textbf{0.7157} & \textbf{0.9373} & \textbf{0.8126} & \textbf{0.7047} & \textbf{0.9486} & \textbf{0.8384} & \textbf{0.7074} & \underline{0.9452} & \textbf{0.8357} & \textbf{0.7143} \\
\bottomrule
\end{tabular}}%
\vspace{-2ex}
\end{table*}

We compare TabTracer with various baselines across three table reasoning datasets (TabFact, WikiTQ, CRT), demonstrating its effectiveness and efficiency.

\subsubsection{Effectiveness evaluation.}
Table~\ref{tab:main_experiment} demonstrates that TabTracer significantly outperforms SOTA baselines, achieving consistently high and stable EM across various datasets and backbones, and ranks at the top in nearly all configurations.
Specifically, on the complex CRT dataset, TabTracer achieves significant improvement, up to 7.4\% over Table-Critic (Qwen3-32B-Direct), with consistent improvements across other LLM backbones. 
Even on the simpler fact-verification dataset (TabFact), TabTracer achieves 0.2–1.7\% improvement, demonstrating the stable performance of TabTracer. 


Prompt-based baselines remain competitive on some settings, especially on fact-style queries. CoT-Consist is the strongest prompt-based baseline in Table~\ref{tab:main_experiment}. It inherits a clear chain-of-thought reasoning path and adds voting across multiple chains, improving robustness. However, it still relies on language-only checks, which hurts numerical tasks.
Agent-based methods invoke tools to reduce model-only hallucinations, and Table-Critic is the strongest agent-based baseline. However, most agents still follow linear trajectories or weak step validation, so errors can propagate, and results vary more across backbones.
TabTracer is best or second-best across datasets and backbones. It performs well because schema-aware subtable slicing reduces distractors, step-level verification catches intermediate errors, and MCTS pruning removes redundant branches while keeping the search focused.


To investigate the impact of reasoning LLMs on table reasoning, we compare representative reasoning LLMs in Table~\ref{tab:main_experiment}. 
It can be seen that existing prompt-based and agent-based methods exhibit strong sensitivity to the LLM backbones. 
For instance, compared with Qwen3-32B, CoT-Consist and Chain-of-Table degrade substantially on Qwen3-32B-Direct, highlighting their dependence on model-level reasoning capabilities. 
In contrast, TabTracer remains robust across model types, consistently achieving the best or second-best performance. 

\begin{table*}[t]
\centering
\caption{Token usage across datasets (Input / Output tokens).}
\vspace{-3ex}
\label{tab:token_usage}
\resizebox{0.95\textwidth}{!}{%
\begin{tabular}{l|cc|cc|cc|cc|cc|cc}
\toprule
\multirow{3}{*}{\textbf{Method}} & \multicolumn{6}{c|}{\textbf{Qwen3-32B-Direct}} & \multicolumn{6}{c}{\textbf{Qwen2.5-72B}} \\
\cmidrule(lr){2-7} \cmidrule(lr){8-13}
 & \multicolumn{2}{c}{\textbf{TabFact}} & \multicolumn{2}{c}{\textbf{WikiTQ}} & \multicolumn{2}{c}{\textbf{CRT}} & \multicolumn{2}{c}{\textbf{TabFact}} & \multicolumn{2}{c}{\textbf{WikiTQ}} & \multicolumn{2}{c}{\textbf{CRT}} \\
 & Input & Output & Input & Output & Input & Output & Input & Output & Input & Output & Input & Output \\
\midrule
CoT-Consist & 3634.02 & 919.18 & \underline{4966.36} & 832.27 & \underline{3494.43} & 1410.83 & 3624.04 & \textbf{839.23} & \underline{4928.08} & 876.33 & \underline{3474.43} & 1535.82 \\
Chain-of-Table & 15212.50 & 1160.23 & 9010.26 & \underline{550.63} & 9587.52 & \textbf{386.70} & 12656.28 & 907.94 & 8665.55 & \textbf{447.27} & 7701.82 & \textbf{370.02} \\
Reactable & 7250.02 & 680.63 & 19091.73 & 663.83 & 8633.00 & 1022.57 & 21530.03 & \underline{864.07} & 21530.03 & 864.07 & 12832.69 & 1337.33 \\
MACT & \textbf{1140.78} & 2232.01 & \textbf{2307.44} & 1391.60 & \textbf{2177.00} & 2188.53 & \textbf{1136.78} & 1708.78 & \textbf{2303.43} & 1371.15 & \textbf{2173.00} & 2044.58 \\
Table-Critic & 9531.60 & \underline{366.20} & 12032.00 & \textbf{528.70} & 13337.60 & 796.30 & 23559.30 & 1713.30 & 29660.70 & 1653.10 & 26571.20 & 1151.90 \\
TabTracer (Ours) & \underline{1983.00} & \textbf{251.40} & 6650.40 & 685.80 & 4373.80 & \underline{533.40} & \underline{2913.50} & 911.70 & 7759.60 & \underline{757.30} & 4926.10 & \underline{940.10} \\
\bottomrule
\end{tabular}}%
\vspace{0.8ex}
\resizebox{0.95\textwidth}{!}{%
\begin{tabular}{l|cc|cc|cc|cc|cc|cc}
\toprule
\multirow{3}{*}{\textbf{Method}} & \multicolumn{6}{c|}{\textbf{Qwen3-32B}} & \multicolumn{6}{c}{\textbf{GPT-4.1-Mini}} \\
\cmidrule(lr){2-7} \cmidrule(lr){8-13}
 & \multicolumn{2}{c}{\textbf{TabFact}} & \multicolumn{2}{c}{\textbf{WikiTQ}} & \multicolumn{2}{c}{\textbf{CRT}} & \multicolumn{2}{c}{\textbf{TabFact}} & \multicolumn{2}{c}{\textbf{WikiTQ}} & \multicolumn{2}{c}{\textbf{CRT}} \\
 & Input & Output & Input & Output & Input & Output & Input & Output & Input & Output & Input & Output \\
\midrule
CoT-Consist & 3715.40 & \textbf{954.12} & \underline{4918.51} & \underline{2323.75} & \underline{3464.43} & 3992.56 & 3157.56 & \underline{710.49} & \underline{4298.59} & \underline{773.76} & \underline{3072.58} & 1357.60 \\
Chain-of-Table & 11677.74 & 8976.12 & 5559.84 & 2338.02 & 7546.68 & 6655.20 & 12880.67 & 1169.83 & 12170.07 & 924.73 & 10806.01 & \textbf{581.89} \\
Reactable & 7537.09 & 3849.19 & 19142.75 & 3939.10 & 8761.14 & 4321.98 & 8696.00 & 2184.04 & 29323.53 & 2153.97 & 13997.54 & 2323.82 \\
MACT & \textbf{1136.78} & 4510.92 & \textbf{2303.40} & 3947.27 & \textbf{2173.00} & 6346.85 & \textbf{1028.51} & 3784.13 & \textbf{1995.81} & 1176.87 & \textbf{1418.05} & 1392.37 \\
Table-Critic & 10337.80 & \underline{1255.80} & 13123.00 & 2609.30 & 15495.50 & \underline{2702.60} & 18378.20 & \textbf{609.00} & 22407.10 & 819.80 & 23457.30 & \underline{1055.10} \\
TabTracer (Ours) & \underline{1984.40} & 1752.20 & 6581.00 & \textbf{1390.40} & 4635.70 & \textbf{1365.50} & \underline{1959.30} & 911.20 & 6831.80 & \textbf{720.00} & 5274.50 & 1182.90 \\
\bottomrule
\end{tabular}}%
\vspace{-2ex}
\end{table*}

\subsubsection{Efficiency evaluation.}

Table~\ref{tab:token_usage} compares token consumption across methods. Prompt-based systems are the most efficient because they run in a single pass or a small number of samples without repeated tool calls. Tool-intensive agents are the most expensive group because iterative tool execution repeatedly reintroduces large table context. Tree-based agents sit between these extremes. Their multi-round critique improves accuracy but adds extra calls and repeated context.
MACT is a special case with low input but high output. The multi-agent dialogue inflates output tokens even when inputs are small.
TabTracer is the most cost-effective among agent systems. Subtable slicing shrinks context early, MCTS pruning removes redundant branches, and the single-step execute-and-reflect loop keeps budgets controlled while preserving accuracy.

Cross-backbone stability analysis shows that token usage can vary widely across LLMs for tool-based agents. The clearest contrast appears between Qwen3-32B and Qwen3-32B-Direct, where the thinking mode produces much longer outputs. Methods that generate long reasoning text at each step amplify this gap because every tool call includes extra model output.
TabTracer shows a smaller variance across backbones. It limits the number of tool calls per query and keeps step outputs short and structured, so backbone verbosity has less room to expand. This indicates that the stability we observe is driven by the method design rather than a specific backbone.



\vspace{-2ex}
\subsection{Numerical Robustness}
\label{subsec:numerical}
Numerical robustness is critical for table reasoning: arithmetic hallucinations easily propagate across chained operations and flip final decisions. Overall TabFact performance already reflects gains in factual faithfulness, and to isolate numerical hallucinations, we extract arithmetic-heavy subsets from WikiTQ and CRT for focused analysis. 
Table~\ref{tab:math_analysis} evaluates TabTracer's ability to suppress numerical hallucinations on arithmetic-intensive subsets of WikiTQ and CRT. 
We analyze TabTracer's performance across three dimensions: task-specific accuracy gains, cross-model stability, and comparison with baseline approaches.
\begin{table}[t]
\centering
\caption{Numerical reasoning accuracy on WikiTQ and CRT (Exact Match). Sample counts: WikiTQ = 2229, CRT = 204.}
\vspace{-2ex}
\label{tab:math_analysis}
{\setlength{\tabcolsep}{3pt}\renewcommand{\arraystretch}{1.05}\resizebox{\columnwidth}{!}{%
\begin{tabular}{l|cc|cc|cc|cc}
\toprule
\multirow{2}{*}{\textbf{Method}} & \multicolumn{2}{c|}{\textbf{Qwen3-32B-Direct}} & \multicolumn{2}{c|}{\textbf{Qwen2.5-72B}} & \multicolumn{2}{c|}{\textbf{Qwen3-32B}} & \multicolumn{2}{c}{\textbf{GPT-4.1-Mini}} \\
 & WikiTQ & CRT & WikiTQ & CRT & WikiTQ & CRT & WikiTQ & CRT \\
\midrule
\multicolumn{9}{c}{\textit{Prompt-Based Reasoning}} \\
\midrule
End-to-End QA & 0.5034 & 0.2108 & 0.5509 & 0.2500 & 0.7856 & 0.4069 & 0.5998 & 0.2921 \\
Few-Shot QA & 0.5334 & 0.2206 & 0.5352 & 0.1961 & 0.7878 & 0.3824 & 0.4773 & 0.1881 \\
Chain-of-Thought & 0.6123 & 0.2549 & 0.7511 & 0.3873 & 0.8054 & 0.4510 & 0.7984 & 0.4406 \\
CoT-Consist & 0.6831 & 0.2843 & \underline{0.7712} & 0.4020 & 0.8228 & 0.4412 & 0.8153 & 0.4505 \\
Lotus & 0.5379 & 0.2451 & 0.6326 & 0.3092 & 0.7672 & 0.3921 & 0.6376 & 0.3333 \\
\midrule
\multicolumn{9}{c}{\textit{Agent-Based Systems}} \\
\midrule
Chain-of-Table & 0.5283 & 0.3465 & 0.6659 & 0.3529 & 0.7798 & 0.4118 & 0.5422 & 0.2637 \\
Reactable & 0.5362 & 0.3333 & 0.4613 & 0.1176 & 0.5344 & 0.3627 & 0.4694 & 0.1485 \\
MACT & 0.6754 & 0.3235 & 0.7001 & 0.3186 & 0.8032 & \underline{0.4608} & 0.6404 & 0.3366 \\
Table-Critic & \underline{0.8362} & \underline{0.4604} & 0.7649 & \underline{0.4149} & \textbf{0.8385} & 0.4406 & \underline{0.8205} & \underline{0.4656} \\
 \textbf{TabTracer (Ours)} & \textbf{0.8398} & \textbf{0.5743} & \textbf{0.7873} & \textbf{0.5347} & \underline{0.8277} & \textbf{0.5644} & \textbf{0.8250} & \textbf{0.5545} \\
\bottomrule
\end{tabular}}%
}
\vspace{-2ex}
\end{table}

Baselines show clear limitations in numerical reasoning. Prompt-based approaches plateau at 45\% on CRT, indicating that longer textual chains still struggle with exact computation. Agent systems without iterative refinement (e.g., Chain-of-Table, MACT) remain below 47\% on CRT, while Table-Critic is the strongest baseline with 41–47\% on CRT and around 76–84\% on WikiTQ arithmetic subsets.
TabTracer substantially outperforms these baselines. On CRT arithmetic queries, it reaches 53–57\% across backbones and improves over Table-Critic by 8.9–12.4 percentage points. On WikiTQ arithmetic subsets, the gains are smaller but consistent at 0.3–2.5 points.
These results directly address numerical hallucination. By isolating arithmetic-heavy subsets, we show that TabTracer reduces numeric hallucinations rather than only improving overall EM.
The gains come from execution-grounded refinement. The tool–reflection loop corrects intermediate arithmetic errors and forces computations to align with table schemas, which is hard to achieve with language-only reasoning or single-pass execution.
TabTracer also shows stronger cross-model stability. Its CRT accuracy stays within a 4\% band (53.5–57.4\%), whereas Table-Critic varies more and stays lower (41.5–46.6\%). This stability indicates that TabTracer’s improvements are driven by the method design rather than a specific backbone.


\begin{table*}[tb]
\centering
\caption{Small/Medium/Large accuracy comparison across datasets and model configurations (Exact Match), and T. denotes TabFact, W. refers to WikiTQ, and C. signifies CRT.}
\vspace{-2ex}
\label{tab:S_M_L}
\resizebox{0.88\textwidth}{!}{%
\begin{tabular}{l|ccc|ccc|ccc|ccc|ccc|ccc}
\toprule
\multirow{3}{*}{\textbf{Method}} & \multicolumn{9}{c|}{\textbf{Qwen3-32B-Direct}} & \multicolumn{9}{c}{\textbf{Qwen2.5-72B}} \\
\cmidrule(lr){2-10} \cmidrule(lr){11-19}
 & \multicolumn{3}{c}{\textbf{Small}} & \multicolumn{3}{c}{\textbf{Medium}} & \multicolumn{3}{c}{\textbf{Large}} & \multicolumn{3}{c}{\textbf{Small}} & \multicolumn{3}{c}{\textbf{Medium}} & \multicolumn{3}{c}{\textbf{Large}} \\
\cmidrule(lr){2-4} \cmidrule(lr){5-7} \cmidrule(lr){8-10} \cmidrule(lr){11-13} \cmidrule(lr){14-16} \cmidrule(lr){17-19}
 & T. & W. & C. & T. & W. & C. & T. & W. & C. & T. & W. & C. & T. & W. & C. & T. & W. & C. \\
\midrule
End-to-End QA & \cellcolor{green!25}0.81 & \cellcolor{green!40}0.62 & \cellcolor{green!40}0.51 & \cellcolor{green!8}0.76 & \cellcolor{green!25}0.52 & \cellcolor{green!25}0.45 & \cellcolor{green!40}0.83 & \cellcolor{green!8}0.45 & \cellcolor{green!8}0.25 & \cellcolor{green!40}0.86 & \cellcolor{green!40}0.68 & \cellcolor{green!40}0.57 & \cellcolor{green!25}0.84 & \cellcolor{green!25}0.60 & \cellcolor{green!25}0.45 & \cellcolor{green!8}0.83 & \cellcolor{green!8}0.49 & \cellcolor{green!8}0.25 \\
Chain-of-Thought & \cellcolor{green!40}0.93 & \cellcolor{green!40}0.60 & \cellcolor{green!25}0.50 & \cellcolor{green!8}0.88 & \cellcolor{green!25}0.51 & \cellcolor{green!8}0.42 & \cellcolor{green!25}0.91 & \cellcolor{green!8}0.37 & \cellcolor{green!40}\textbf{0.62} & \cellcolor{green!25}0.93 & \cellcolor{green!40}0.81 & \cellcolor{green!40}0.66 & \cellcolor{green!8}0.91 & \cellcolor{green!25}0.76 & \cellcolor{green!25}0.55 & \cellcolor{green!40}\underline{0.96} & \cellcolor{green!8}0.61 & \cellcolor{green!8}\underline{0.50} \\
CoT-Consist & \cellcolor{green!25}0.94 & \cellcolor{green!40}0.65 & \cellcolor{green!25}0.53 & \cellcolor{green!8}0.92 & \cellcolor{green!25}0.57 & \cellcolor{green!8}0.43 & \cellcolor{green!40}\underline{0.96} & \cellcolor{green!8}0.42 & \cellcolor{green!40}\textbf{0.62} & \cellcolor{green!25}\textbf{0.94} & \cellcolor{green!40}0.82 & \cellcolor{green!40}0.66 & \cellcolor{green!8}0.93 & \cellcolor{green!25}\underline{0.78} & \cellcolor{green!25}\underline{0.58} & \cellcolor{green!40}\textbf{1.00} & \cellcolor{green!8}\underline{0.65} & \cellcolor{green!8}\underline{0.50} \\
\midrule
Chain-of-Table & \cellcolor{green!25}0.90 & \cellcolor{green!40}0.53 & \cellcolor{green!40}0.55 & \cellcolor{green!8}0.86 & \cellcolor{green!25}0.52 & \cellcolor{green!25}0.45 & \cellcolor{green!40}\textbf{1.00} & \cellcolor{green!8}0.45 & \cellcolor{green!8}0.25 & \cellcolor{green!25}0.91 & \cellcolor{green!40}0.76 & \cellcolor{green!40}0.60 & \cellcolor{green!8}0.89 & \cellcolor{green!25}0.70 & \cellcolor{green!25}0.48 & \cellcolor{green!40}\textbf{1.00} & \cellcolor{green!8}0.56 & \cellcolor{green!8}0.38 \\
MACT & \cellcolor{green!25}0.88 & \cellcolor{green!40}0.76 & \cellcolor{green!40}0.61 & \cellcolor{green!8}0.85 & \cellcolor{green!25}0.69 & \cellcolor{green!25}0.53 & \cellcolor{green!40}\underline{0.96} & \cellcolor{green!8}0.61 & \cellcolor{green!8}\underline{0.50} & \cellcolor{green!40}0.91 & \cellcolor{green!40}0.78 & \cellcolor{green!40}0.62 & \cellcolor{green!25}0.87 & \cellcolor{green!25}0.70 & \cellcolor{green!25}0.54 & \cellcolor{green!25}0.87 & \cellcolor{green!8}0.63 & \cellcolor{green!8}\underline{0.50} \\
Table-Critic & \cellcolor{green!8}\underline{0.94} & \cellcolor{green!40}\underline{0.86} & \cellcolor{green!40}\underline{0.68} & \cellcolor{green!25}\textbf{0.95} & \cellcolor{green!25}\underline{0.84} & \cellcolor{green!25}\underline{0.56} & \cellcolor{green!40}\underline{0.96} & \cellcolor{green!8}\underline{0.68} & \cellcolor{green!8}\underline{0.50} & \cellcolor{green!25}0.93 & \cellcolor{green!40}\underline{0.82} & \cellcolor{green!40}\underline{0.66} & \cellcolor{green!25}\underline{0.93} & \cellcolor{green!25}0.76 & \cellcolor{green!25}0.54 & \cellcolor{green!40}\underline{0.96} & \cellcolor{green!8}0.64 & \cellcolor{green!8}\underline{0.50} \\
\textbf{TabTracer (Ours)} & \cellcolor{green!25}\textbf{0.94} & \cellcolor{green!40}\textbf{0.87} & \cellcolor{green!40}\textbf{0.72} & \cellcolor{green!25}\underline{0.94} & \cellcolor{green!25}\textbf{0.84} & \cellcolor{green!25}\textbf{0.70} & \cellcolor{green!40}\textbf{1.00} & \cellcolor{green!8}\textbf{0.79} & \cellcolor{green!8}\textbf{0.62} & \cellcolor{green!25}\underline{0.94} & \cellcolor{green!40}\textbf{0.84} & \cellcolor{green!40}\textbf{0.72} & \cellcolor{green!8}\textbf{0.93} & \cellcolor{green!25}\textbf{0.81} & \cellcolor{green!25}\textbf{0.67} & \cellcolor{green!40}\textbf{1.00} & \cellcolor{green!8}\textbf{0.73} & \cellcolor{green!8}\textbf{0.62} \\
\bottomrule
\end{tabular}}%
\vspace{0.8ex}
\resizebox{0.88\textwidth}{!}{%
\begin{tabular}{l|ccc|ccc|ccc|ccc|ccc|ccc}
\toprule
\multirow{3}{*}{\textbf{Method}} & \multicolumn{9}{c|}{\textbf{Qwen3-32B}} & \multicolumn{9}{c}{\textbf{GPT-4.1-Mini}} \\
\cmidrule(lr){2-10} \cmidrule(lr){11-19}
 & \multicolumn{3}{c}{\textbf{Small}} & \multicolumn{3}{c}{\textbf{Medium}} & \multicolumn{3}{c}{\textbf{Large}} & \multicolumn{3}{c}{\textbf{Small}} & \multicolumn{3}{c}{\textbf{Medium}} & \multicolumn{3}{c}{\textbf{Large}} \\
\cmidrule(lr){2-4} \cmidrule(lr){5-7} \cmidrule(lr){8-10} \cmidrule(lr){11-13} \cmidrule(lr){14-16} \cmidrule(lr){17-19}
 & T. & W. & C. & T. & W. & C. & T. & W. & C. & T. & W. & C. & T. & W. & C. & T. & W. & C. \\
\midrule
End-to-End QA & \cellcolor{green!25}\textbf{0.95} & \cellcolor{green!40}0.81 & \cellcolor{green!40}0.62 & \cellcolor{green!8}0.93 & \cellcolor{green!25}0.75 & \cellcolor{green!8}0.57 & \cellcolor{green!40}\underline{0.96} & \cellcolor{green!8}0.61 & \cellcolor{green!40}\underline{0.62} & \cellcolor{green!40}0.87 & \cellcolor{green!40}0.72 & \cellcolor{green!40}0.57 & \cellcolor{green!8}0.84 & \cellcolor{green!25}0.64 & \cellcolor{green!25}0.52 & \cellcolor{green!40}0.87 & \cellcolor{green!8}0.51 & \cellcolor{green!8}0.38 \\
Chain-of-Thought & \cellcolor{green!25}0.93 & \cellcolor{green!40}0.78 & \cellcolor{green!25}0.64 & \cellcolor{green!25}0.93 & \cellcolor{green!25}0.76 & \cellcolor{green!8}0.57 & \cellcolor{green!40}\textbf{1.00} & \cellcolor{green!8}0.61 & \cellcolor{green!40}\textbf{0.75} & \cellcolor{green!25}0.94 & \cellcolor{green!40}0.79 & \cellcolor{green!40}0.63 & \cellcolor{green!25}0.94 & \cellcolor{green!25}0.75 & \cellcolor{green!8}0.53 & \cellcolor{green!40}\textbf{1.00} & \cellcolor{green!8}0.62 & \cellcolor{green!25}\textbf{0.62} \\
CoT-Consist & \cellcolor{green!8}0.94 & \cellcolor{green!40}0.80 & \cellcolor{green!25}0.63 & \cellcolor{green!25}\underline{0.95} & \cellcolor{green!25}0.79 & \cellcolor{green!8}0.58 & \cellcolor{green!40}\textbf{1.00} & \cellcolor{green!8}0.64 & \cellcolor{green!40}\textbf{0.75} & \cellcolor{green!25}\textbf{0.95} & \cellcolor{green!40}0.80 & \cellcolor{green!40}0.63 & \cellcolor{green!25}\textbf{0.95} & \cellcolor{green!25}0.78 & \cellcolor{green!8}0.54 & \cellcolor{green!40}\textbf{1.00} & \cellcolor{green!8}0.63 & \cellcolor{green!25}\textbf{0.62} \\
\midrule
Chain-of-Table & \cellcolor{green!40}0.92 & \cellcolor{green!40}0.77 & \cellcolor{green!40}0.64 & \cellcolor{green!8}0.89 & \cellcolor{green!25}0.76 & \cellcolor{green!25}0.57 & \cellcolor{green!25}0.91 & \cellcolor{green!8}0.64 & \cellcolor{green!8}0.50 & \cellcolor{green!25}0.87 & \cellcolor{green!40}0.65 & \cellcolor{green!25}0.59 & \cellcolor{green!8}0.86 & \cellcolor{green!25}0.61 & \cellcolor{green!8}0.45 & \cellcolor{green!40}\underline{0.95} & \cellcolor{green!8}0.56 & \cellcolor{green!40}\textbf{0.62} \\
MACT & \cellcolor{green!8}0.93 & \cellcolor{green!40}0.82 & \cellcolor{green!40}\underline{0.68} & \cellcolor{green!40}\textbf{0.96} & \cellcolor{green!25}0.79 & \cellcolor{green!8}\underline{0.61} & \cellcolor{green!40}\underline{0.96} & \cellcolor{green!8}0.67 & \cellcolor{green!25}\underline{0.62} & \cellcolor{green!25}0.90 & \cellcolor{green!40}0.72 & \cellcolor{green!40}0.61 & \cellcolor{green!8}0.87 & \cellcolor{green!25}0.64 & \cellcolor{green!25}0.51 & \cellcolor{green!40}\textbf{1.00} & \cellcolor{green!8}0.57 & \cellcolor{green!8}\underline{0.50} \\
Table-Critic & \cellcolor{green!25}0.94 & \cellcolor{green!40}\underline{0.86} & \cellcolor{green!40}0.67 & \cellcolor{green!25}0.94 & \cellcolor{green!25}\textbf{0.84} & \cellcolor{green!25}0.60 & \cellcolor{green!40}\underline{0.96} & \cellcolor{green!8}\underline{0.69} & \cellcolor{green!8}0.50 & \cellcolor{green!25}0.94 & \cellcolor{green!40}\underline{0.84} & \cellcolor{green!40}\underline{0.70} & \cellcolor{green!25}0.94 & \cellcolor{green!25}\underline{0.82} & \cellcolor{green!8}\underline{0.57} & \cellcolor{green!40}\textbf{1.00} & \cellcolor{green!8}\underline{0.68} & \cellcolor{green!25}\textbf{0.62} \\
\textbf{TabTracer (Ours)} & \cellcolor{green!25}\underline{0.95} & \cellcolor{green!40}\textbf{0.86} & \cellcolor{green!40}\textbf{0.72} & \cellcolor{green!8}0.94 & \cellcolor{green!25}\underline{0.83} & \cellcolor{green!25}\textbf{0.67} & \cellcolor{green!40}\textbf{1.00} & \cellcolor{green!8}\textbf{0.77} & \cellcolor{green!8}0.50 & \cellcolor{green!8}\underline{0.94} & \cellcolor{green!40}\textbf{0.86} & \cellcolor{green!40}\textbf{0.73} & \cellcolor{green!25}\underline{0.95} & \cellcolor{green!25}\textbf{0.83} & \cellcolor{green!25}\textbf{0.68} & \cellcolor{green!40}\textbf{1.00} & \cellcolor{green!8}\textbf{0.75} & \cellcolor{green!8}\underline{0.50} \\
\bottomrule
\end{tabular}}%
\vspace{-2ex}
\end{table*}

\vspace{-3ex}
\subsection{Scalability Experiment}
\label{subsec:scalability}
This section examines how TabTracer scales with table sizes.
Table~\ref{tab:S_M_L} stratifies queries by table size (small/medium/large) and uses a three-level green gradient to highlight scaling trends across datasets.

On small tables, TabTracer achieves 94-95\% accuracy on TabFact, reaches 86\% on WikiTQ (matching the leading baseline Table-Critic), and maintains 3-6\% advantages on CRT.
This shows that TabTracer preserves strong accuracy when tables are compact and the evidence is easy to isolate.

As complexity increases to medium tables, TabTracer maintains stable performance around 84\% on WikiTQ while competing methods suffer substantial degradation (CoT-Consist drops 15 points, MACT falls from 82\% to 67\%); on CRT, TabTracer sustains 67-70\% accuracy while Chain-of-Table experiences 12\% drops.
This demonstrates TabTracer's superior resilience compared to single-trajectory approaches, as the tree search mechanism enables exploring multiple paths to handle increased complexity.

On the large tables, TabTracer maintains 73-79\% accuracy on WikiTQ, preserving 7-12 point leads over Table-Critic's 64-68\%, and sustains 12\% advantages on CRT across open-weight models.
Despite natural performance degradation, TabTracer's tree search and reflection mechanisms demonstrate significant advantages in high-complexity scenarios.

Table~\ref{tab:S_M_L} uses a three-level green gradient, where darker cells indicate better accuracy on smaller tables and lighter cells indicate worse accuracy on larger tables. The color trend shows a consistent downward shift from small to large tables across methods, confirming that larger tables are harder. The decline is most visible on WikiTQ and CRT, while TabFact shows a milder pattern because it is a binary verification task. We also note that the large-table subsets for TabFact and CRT are relatively small, so their large-size results can be noisier.
Compared with baselines, TabTracer shows a smaller drop from small to large tables on WikiTQ and CRT, while prompt-based and single-trajectory agents degrade more sharply. This indicates stronger scalability for TabTracer under increasing table size.

\begin{figure}[tbh]
    \centering
    \vspace{7ex}
    \includegraphics[width=0.98\linewidth]{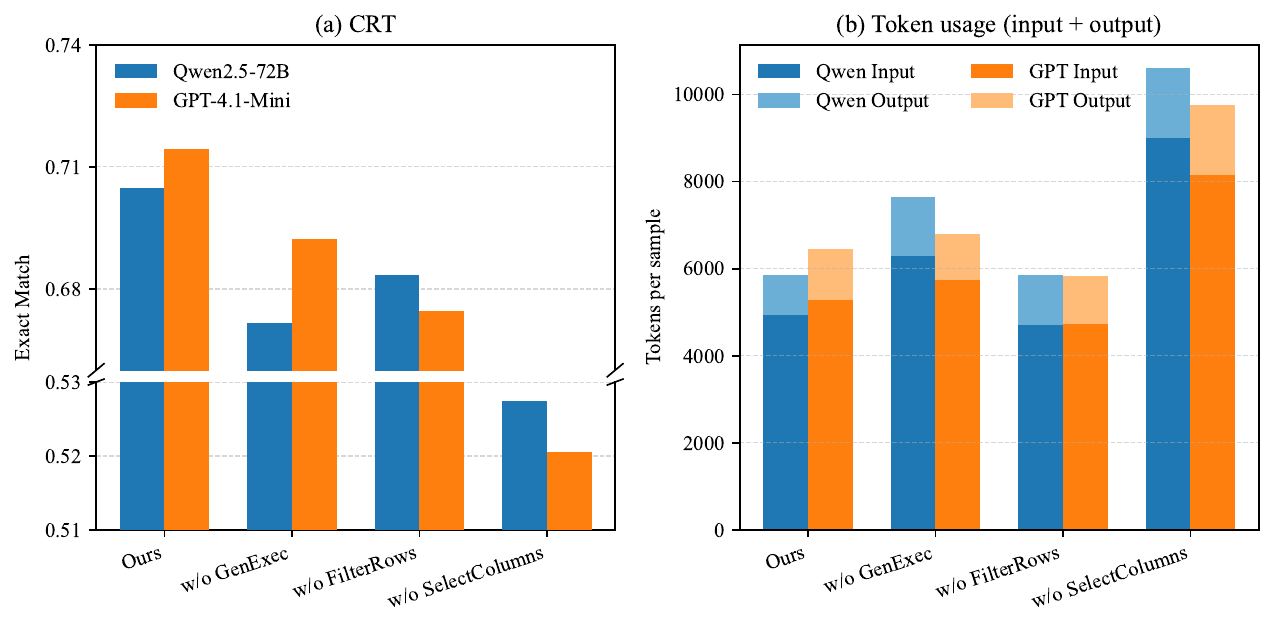}
    \vspace{-3ex}
    \caption{Tool ablation on CRT. We report exact-match accuracy and token usage when removing each operator.}
    \vspace{-2ex}
    \label{fig:tool_ablation}
\end{figure}
\subsection{Ablation Study}
\label{subsec:ablation}
We validate the necessity of each tool through single-component ablation experiments on CRT, measuring both accuracy and token efficiency impacts as shown in Figure~\ref{fig:tool_ablation}. Removing any tool reduces exact-match accuracy. Token usage also rises when the planner cannot delegate. The largest increase appears without SelectColumns, and dropping GenExecCode adds around $1$k tokens despite lower accuracy.

Removing \texttt{SelectColumns} causes the most severe degradation: exact-match accuracy plummets from $70.5\%$ to $52$--$53\%$ (a $17$--$19$ percentage point drop), representing the largest accuracy impact among all three tools.
Simultaneously, input token consumption surges from $4.9k$ to $9.0k$ tokens for Qwen2.5-72B, nearly doubling the computational budget.
This dual impact stems from \texttt{SelectColumns}' role in table width reduction—without column pruning, every subsequent tool invocation carries the full table width, inflating context length while forcing the model to reason over irrelevant columns.
SelectColumns thus emerges as the critical component for both accuracy and efficiency.

Removing \texttt{FilterRows} reduces accuracy by $2$--$4$ percentage points while increasing token consumption by approximately $600$ tokens.
FilterRows enables row-level filtering to eliminate distractor information, and its absence forces the system to search through larger row spaces, increasing reasoning difficulty.
Although the individual impact is smaller than SelectColumns, the cumulative effect across multi-step reasoning chains remains significant.

Removing \texttt{GenExecCode} decreases accuracy by $2$--$4$ percentage points, yet paradoxically increases token consumption by approximately $1k$ tokens.
This counterintuitive pattern arises because, without code generation capabilities, the system resorts to lengthier natural language descriptions and multiple tool invocations to accomplish complex computations, resulting in token inflation despite reduced accuracy.
GenExecCode therefore provides a concise and efficient computational pathway that outperforms natural-language alternatives.


The complete toolset achieves optimal performance in both accuracy and efficiency dimensions.
Any single tool's removal either degrades accuracy or increases token consumption, validating the necessity of each tool in the architecture.
\subsection{Component Analysis}
\label{subsec:component}

This analysis probes step-level reliability and search behavior, where error propagation can arise without intermediate verification. Therefore, we report tool success/adoption and search diagnostics.

\begin{figure*}[t]
  \centering
  \includegraphics[width=\textwidth]{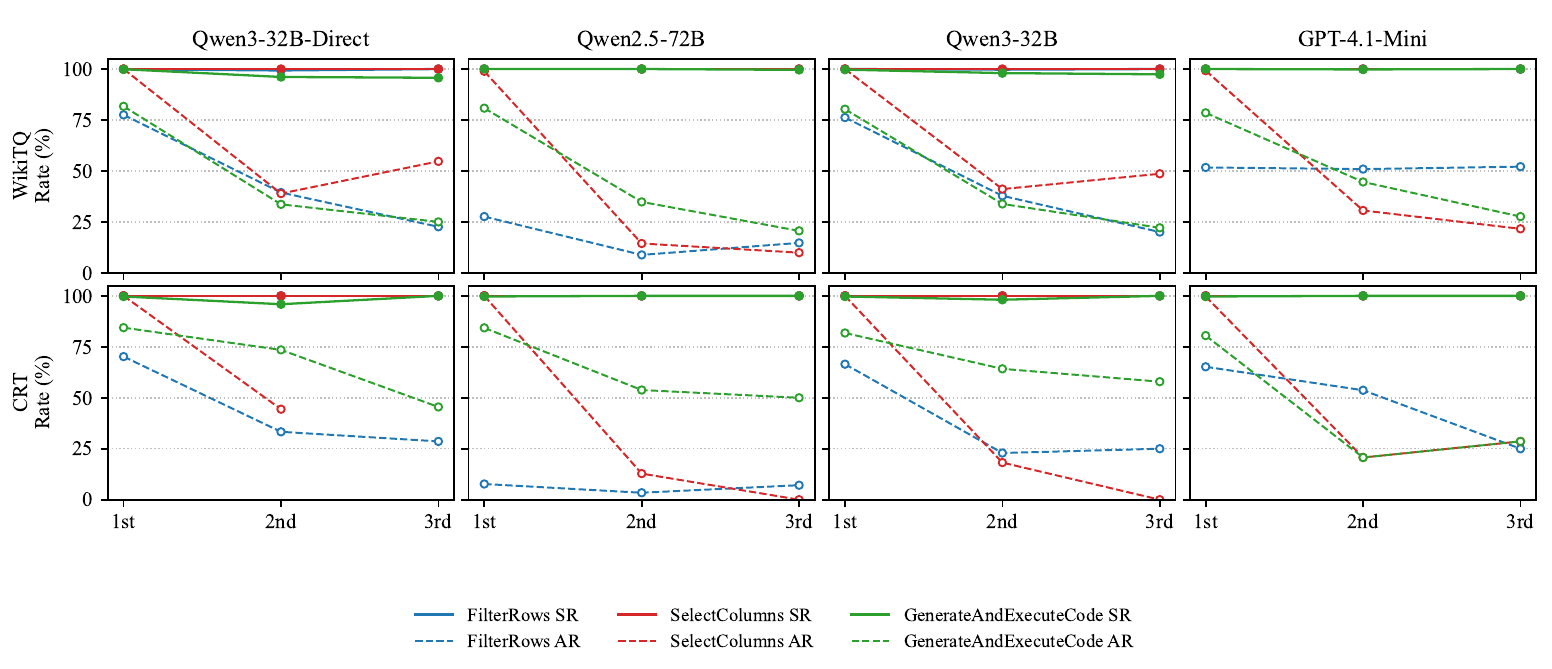}
  \caption{Tool success rate (SR) and adoption rate (AR) across stages.}
  \label{fig:system_performance}
\end{figure*}

\begin{figure}[tb]
    \centering
    \vspace{4ex}
    \includegraphics[width=\linewidth]{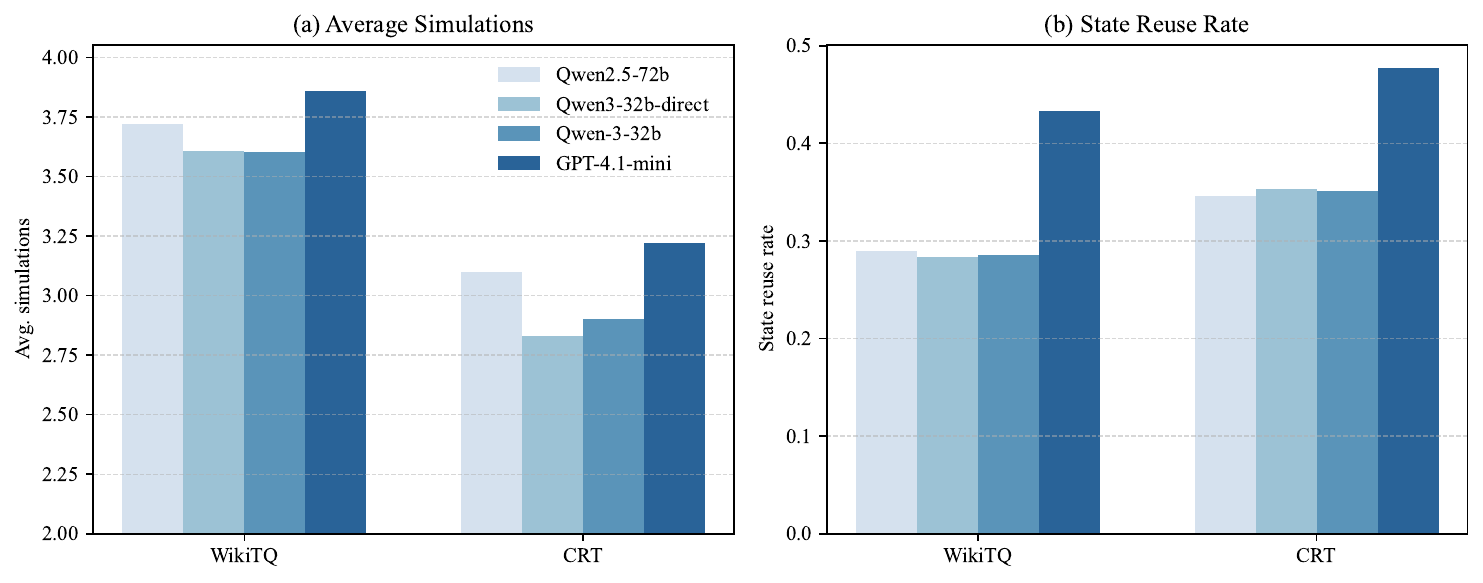}
    \vspace{-4ex}
    \caption{Average simulations and state reuse rate on WikiTQ and CRT.}
    \vspace{-5ex}
    \label{fig:state_reuse}
\end{figure}

\begin{figure}[tbh]
    \centering
    \vspace{2ex}
    \includegraphics[width=\linewidth]{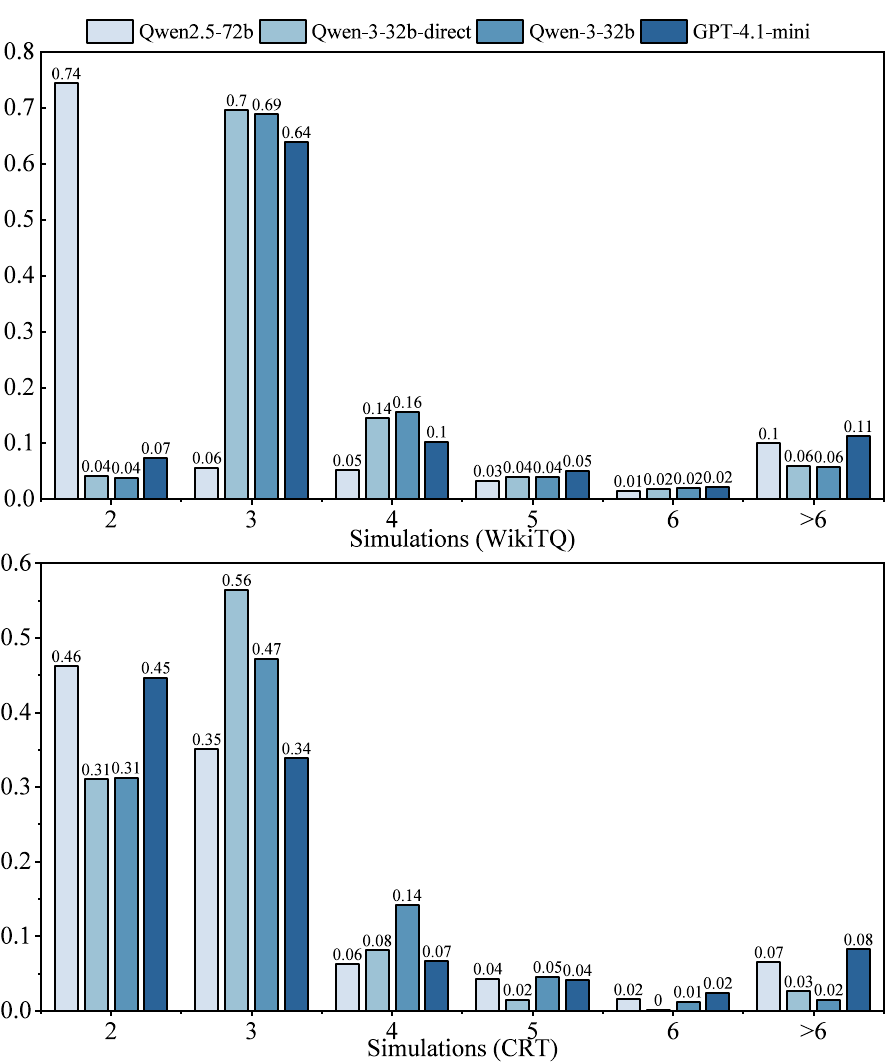}
    \vspace{-5ex}
    \caption{Distribution of simulations on WikiTQ and CRT.}
    \vspace{-5ex}
    \label{fig:simulation_counts}
\end{figure}

Figure~\ref{fig:system_performance}-\ref{fig:state_reuse}-\ref{fig:simulation_counts} 
underpin this component analysis: we first examine operational stability (tool success and adoption), then inspect search behaviour (state reuse and simulation counts). 

Figure~\ref{fig:system_performance} shows that every tool clears the first-attempt success rate above $99.7\%$ across models. FilterRows and SelectColumns hit $99.9$--$100\%$, indicating that even complex filters execute deterministically, while GenExecCode maintains $99.7$--$100\%$ despite synthesising bespoke Python code. 

First-attempt adoption mirrors task difficulty. 
FilterRows retains $77.5\%$ of outputs on WikiTQ with Qwen3-32B-Direct and $70.2\%$ on CRT, but drops to $27.6\%$ on WikiTQ with Qwen2.5-72B as the model explores broader branches. 
Subsequent attempts are essential: the second invocation recovers an additional 8.8\% of answers and the third a further 14.7\%. 
Conversely, GPT-4.1-mini starts higher on CRT (65.2\%) and leverages a strong second attempt (53.7\%) to close residual gaps. 
SelectColumns remains nearly perfect ($\geq$98.9\% first-attempt adoption), confirming column selection as a solved subtask. 


Our proposed MCTS-based architecture demonstrates strong operational efficiency, validating the choice of tree-guided exploration over exhaustive reasoning.
Figure~\ref{fig:state_reuse} demonstrates that TabTracer maintains modest simulation counts (2.83–3.86 per query) while reusing up to 47.7\% of cached states, providing evidence that the caching strategy effectively amortizes search cost. 
As shown in Figure~\ref{fig:simulation_counts}, 64–88\% of queries converge within three simulations, while extensive searches (>6 expansions) remain below 11\%, illustrating well-controlled exploration.
These results validate that search architecture achieves a balance between exploration depth and computational efficiency. This happens because reflection-based scores guide selection and hash-based filtering removes repeated state-action pairs, which prunes redundant branches early.

\section{Conclusion}
\label{sec:conclusion}
In this paper, we introduce TabTracer, a MCTS agent that restores discipline to complex table reasoning. We first decouple planning from execution: an information-augmented UCB1 policy expands tool calls only when an intermediate question promises new evidence, while immediate reflection scores each branch without long rollouts. We then ground reasoning in deterministic dataframe tools, yielding auditable operations that suppress hallucinations. Finally, TabTracer persists versioned table states to enable low-cost backtracking and state reuse, keeping exploration budget in check. 



\bibliographystyle{ACM-Reference-Format}
\bibliography{tabtracer}

@article{gorishniy2021revisiting,
  title={Revisiting deep learning models for tabular data},
  author={Gorishniy, Yury and Rubachev, Ivan and Khrulkov, Valentin and Babenko, Artem},
  journal={Advances in neural information processing systems},
  volume={34},
  pages={18932--18943},
  year={2021}
}

@inproceedings{borisovlanguage,
  title={Language Models are Realistic Tabular Data Generators},
  author={Borisov, Vadim and Sessler, Kathrin and Leemann, Tobias and Pawelczyk, Martin and Kasneci, Gjergji},
  booktitle={The Eleventh International Conference on Learning Representations}
}

@article{cai2024cohortnet,
  title={CohortNet: Empowering Cohort Discovery for Interpretable Healthcare Analytics},
  author={Cai, Qingpeng and Zheng, Kaiping and Jagadish, HV and Ooi, Beng Chin and Yip, James},
  journal={Proceedings of the VLDB Endowment},
  volume={17},
  number={10},
  pages={2487--2500},
  year={2024},
  publisher={VLDB Endowment}
}

@article{altman2023realistic,
  title={Realistic synthetic financial transactions for anti-money laundering models},
  author={Altman, Erik and Blanu{\v{s}}a, Jovan and Von Niederh{\"a}usern, Luc and Egressy, B{\'e}ni and Anghel, Andreea and Atasu, Kubilay},
  journal={Advances in Neural Information Processing Systems},
  volume={36},
  pages={29851--29874},
  year={2023}
}

@article{burdick2020table,
  title={Table extraction and understanding for scientific and enterprise applications},
  author={Burdick, Douglas and Danilevsky, Marina and Evfimievski, Alexandre V and Katsis, Yannis and Wang, Nancy},
  journal={Proceedings of the VLDB Endowment},
  volume={13},
  number={12},
  pages={3433--3436},
  year={2020},
  publisher={VLDB Endowment}
}

@article{fanglarge,
  title={Large Language Models (LLMs) on Tabular Data: Prediction, Generation, and Understanding-A Survey},
  author={Fang, Xi and Xu, Weijie and Tan, Fiona Anting and Hu, Ziqing and Zhang, Jiani and Qi, Yanjun and Sengamedu, Srinivasan H and Faloutsos, Christos},
  journal={Transactions on Machine Learning Research}
}

@inproceedings{pasupat2015compositional,
  title={Compositional Semantic Parsing on Semi-Structured Tables},
  author={Pasupat, Panupong and Liang, Percy},
  booktitle={Proceedings of the 53rd Annual Meeting of the Association for Computational Linguistics and the 7th International Joint Conference on Natural Language Processing (Volume 1: Long Papers)},
  pages={1470--1480},
  year={2015}
}

@inproceedings{pasupat2016inferring,
  title={Inferring Logical Forms From Denotations},
  author={Pasupat, Panupong and Liang, Percy},
  booktitle={Proceedings of the 54th Annual Meeting of the Association for Computational Linguistics (Volume 1: Long Papers)},
  pages={23--32},
  year={2016}
}

@inproceedings{zhang2017macro,
  title={Macro Grammars and Holistic Triggering for Efficient Semantic Parsing},
  author={Zhang, Yuchen and Pasupat, Panupong and Liang, Percy},
  booktitle={Proceedings of the 2017 Conference on Empirical Methods in Natural Language Processing},
  pages={1214--1223},
  year={2017}
}

@inproceedings{yin2020tabert,
  title={TaBERT: Pretraining for Joint Understanding of Textual and Tabular Data},
  author={Yin, Pengcheng and Neubig, Graham and Yih, Wen-tau and Riedel, Sebastian},
  booktitle={Proceedings of the 58th Annual Meeting of the Association for Computational Linguistics},
  pages={8413--8426},
  year={2020}
}

@inproceedings{herzig2020tapas,
  title={TaPas: Weakly supervised table parsing via pre-training},
  author={Herzig, Jonathan and Nowak, Pawel Krzysztof and M{\"u}ller, Thomas and Piccinno, Francesco and Eisenschlos, Julian},
  booktitle={Proceedings of the 58th annual meeting of the association for computational linguistics},
  pages={4320--4333},
  year={2020}
}

@inproceedings{jauhar2016tables,
  title={Tables as semi-structured knowledge for question answering},
  author={Jauhar, Sujay Kumar and Turney, Peter and Hovy, Eduard},
  booktitle={Proceedings of the 54th Annual Meeting of the Association for Computational Linguistics (Volume 1: Long Papers)},
  pages={474--483},
  year={2016}
}

@article{yang2025qwen3,
  title={Qwen3 technical report},
  author={Yang, An and Li, Anfeng and Yang, Baosong and Zhang, Beichen and Hui, Binyuan and Zheng, Bo and Yu, Bowen and Gao, Chang and Huang, Chengen and Lv, Chenxu and others},
  journal={arXiv preprint arXiv:2505.09388},
  year={2025}
}

@article{hurst2024gpt,
  title={Gpt-4o system card},
  author={Hurst, Aaron and Lerer, Adam and Goucher, Adam P and Perelman, Adam and Ramesh, Aditya and Clark, Aidan and Ostrow, AJ and Welihinda, Akila and Hayes, Alan and Radford, Alec and others},
  journal={arXiv preprint arXiv:2410.21276},
  year={2024}
}

@inproceedings{guo2024large,
  title={Large language model based multi-agents: a survey of progress and challenges},
  author={Guo, Taicheng and Chen, Xiuying and Wang, Yaqi and Chang, Ruidi and Pei, Shichao and Chawla, Nitesh V and Wiest, Olaf and Zhang, Xiangliang},
  booktitle={Proceedings of the Thirty-Third International Joint Conference on Artificial Intelligence},
  pages={8048--8057},
  year={2024}
}

@inproceedings{shankar2024building,
  title={Building reactive large language model pipelines with motion},
  author={Shankar, Shreya and Parameswaran, Aditya G},
  booktitle={Companion of the 2024 International Conference on Management of Data},
  pages={520--523},
  year={2024}
}

@inproceedings{yaoreact,
  title={ReAct: Synergizing Reasoning and Acting in Language Models},
  author={Yao, Shunyu and Zhao, Jeffrey and Yu, Dian and Du, Nan and Shafran, Izhak and Narasimhan, Karthik R and Cao, Yuan},
  booktitle={The Eleventh International Conference on Learning Representations}
}

@article{puttaswamy2025delta,
  title={Delta Sharing: An Open Protocol for Cross-Platform Data Sharing},
  author={Puttaswamy, Krishna and Chakankar, Abhijit and Tao, Tao and Valani, Zaheera and Chandra, Ramesh and Chau, William and Chen, Mengxi and Chetibi, Akram and Huang, Tianyi and Keller, Jonathan and others},
  journal={Proceedings of the VLDB Endowment},
  volume={18},
  number={12},
  pages={5197--5209},
  year={2025},
  publisher={VLDB Endowment}
}

@article{shinn2023reflexion,
  title={Reflexion: Language agents with verbal reinforcement learning},
  author={Shinn, Noah and Cassano, Federico and Gopinath, Ashwin and Narasimhan, Karthik and Yao, Shunyu},
  journal={Advances in Neural Information Processing Systems},
  volume={36},
  pages={8634--8652},
  year={2023}
}

@article{zhang2024chain,
  title={Chain of agents: Large language models collaborating on long-context tasks},
  author={Zhang, Yusen and Sun, Ruoxi and Chen, Yanfei and Pfister, Tomas and Zhang, Rui and Arik, Sercan},
  journal={Advances in Neural Information Processing Systems},
  volume={37},
  pages={132208--132237},
  year={2024}
}

@inproceedings{liuagentbench,
  title={AgentBench: Evaluating LLMs as Agents},
  author={Liu, Xiao and Yu, Hao and Zhang, Hanchen and Xu, Yifan and Lei, Xuanyu and Lai, Hanyu and Gu, Yu and Ding, Hangliang and Men, Kaiwen and Yang, Kejuan and others},
  booktitle={The Twelfth International Conference on Learning Representations}
}

@article{li2024table,
  title={Table-gpt: Table fine-tuned gpt for diverse table tasks},
  author={Li, Peng and He, Yeye and Yashar, Dror and Cui, Weiwei and Ge, Song and Zhang, Haidong and Rifinski Fainman, Danielle and Zhang, Dongmei and Chaudhuri, Surajit},
  journal={Proceedings of the ACM on Management of Data},
  volume={2},
  number={3},
  pages={1--28},
  year={2024},
  publisher={ACM New York, NY, USA}
}

@inproceedings{zhuang2024toolchain,
  title={ToolChain*: Efficient Action Space Navigation in Large Language Models with A* Search},
  author={Zhuang, Yuchen and Chen, Xiang and Yu, Tong and Mitra, Saayan and Bursztyn, Victor and Rossi, Ryan A and Sarkhel, Somdeb and Zhang, Chao},
  booktitle={ICLR},
  year={2024}
}

@article{papotti2025panel,
  title={Panel on Neural Relational Data: Tabular Foundation Models, LLMs... or Both?},
  author={Papotti, Paolo and Binnig, Carsten},
  journal={Proceedings of the VLDB Endowment},
  volume={18},
  number={12},
  pages={5513--5515},
  year={2025},
  publisher={VLDB Endowment}
}

@inproceedings{wan2024alphazero,
  title={AlphaZero-like tree-search can guide large language model decoding and training},
  author={Wan, Ziyu and Feng, Xidong and Wen, Muning and McAleer, Stephen Marcus and Wen, Ying and Zhang, Weinan and Wang, Jun},
  booktitle={Proceedings of the 41st International Conference on Machine Learning},
  pages={49890--49920},
  year={2024}
}

@inproceedings{nahid2024tabsqlify,
  title={TabSQLify: Enhancing Reasoning Capabilities of LLMs Through Table Decomposition},
  author={Nahid, Md and Rafiei, Davood},
  booktitle={Proceedings of the 2024 Conference of the North American Chapter of the Association for Computational Linguistics: Human Language Technologies (Volume 1: Long Papers)},
  pages={5725--5737},
  year={2024}
}

@inproceedings{abhyankar2025h,
  title={H-star: Llm-driven hybrid sql-text adaptive reasoning on tables},
  author={Abhyankar, Nikhil and Gupta, Vivek and Roth, Dan and Reddy, Chandan K},
  booktitle={Proceedings of the 2025 Conference of the Nations of the Americas Chapter of the Association for Computational Linguistics: Human Language Technologies (Volume 1: Long Papers)},
  pages={8841--8863},
  year={2025}
}

@article{jiang2025tablemind,
  title={Tablemind: An autonomous programmatic agent for tool-augmented table reasoning},
  author={Jiang, Chuang and Cheng, Mingyue and Tao, Xiaoyu and Mao, Qingyang and Ouyang, Jie and Liu, Qi},
  journal={arXiv preprint arXiv:2509.06278},
  year={2025}
}

@article{zhu2024autotqa,
  title={Autotqa: Towards autonomous tabular question answering through multi-agent large language models},
  author={Zhu, Jun-Peng and Cai, Peng and Xu, Kai and Li, Li and Sun, Yishen and Zhou, Shuai and Su, Haihuang and Tang, Liu and Liu, Qi},
  journal={Proceedings of the VLDB Endowment},
  volume={17},
  number={12},
  pages={3920--3933},
  year={2024},
  publisher={VLDB Endowment}
}

@article{guo2026rethinking,
  title={Rethinking Table Pruning in TableQA: From Sequential Revisions to Gold Trajectory-Supervised Parallel Search},
  author={Guo, Yu and Ye, Shenghao and Chen, Shuangwu and Wen, Zijian and Zhang, Tao and Bai, Qirui and Jin, Dong and Hou, Yunpeng and He, Huasen and Yang, Jian and others},
  journal={arXiv preprint arXiv:2601.03851},
  year={2026}
}

@inproceedings{jin2025talon,
  title={TALON: A Multi-Agent Framework for Long-Table Exploration and Question Answering},
  author={Jin, Ruochun and Wang, Xiyue and Wang, Dong and Zheng, Haoqi and Qi, Yunpeng and Yang, Silin and Zhang, Meng},
  booktitle={Proceedings of the 2025 Conference on Empirical Methods in Natural Language Processing},
  pages={27385--27401},
  year={2025}
}

@article{zhang2025aixelask,
  title={AixelAsk: A Stepwise-Guided Retrieval and Reasoning Framework for Large Table QA},
  author={Zhang, Chi and Zhang, Meihui and Yang, Yuxin and Chen, Tao and Luo, Zhaojing},
  journal={Proceedings of the ACM on Management of Data},
  volume={3},
  number={6},
  pages={1--25},
  year={2025},
  publisher={ACM New York, NY, USA}
}

@inproceedings{chen2023large,
  title={Large language models are few (1)-shot table reasoners},
  author={Chen, Wenhu},
  booktitle={Findings of the association for computational linguistics: EACL 2023},
  pages={1120--1130},
  year={2023}
}

@article{zhang2024flextaf,
  title={Flextaf: Enhancing table reasoning with flexible tabular formats},
  author={Zhang, Xuanliang and Wang, Dingzirui and Dou, Longxu and Wang, Baoxin and Wu, Dayong and Zhu, Qingfu and Che, Wanxiang},
  journal={arXiv preprint arXiv:2408.08841},
  year={2024}
}

@inproceedings{ziqi2023tab,
  title={Tab-cot: Zero-shot tabular chain of thought},
  author={Ziqi, Jin and Lu, Wei},
  booktitle={Findings of the Association for Computational Linguistics: ACL 2023},
  pages={10259--10277},
  year={2023}
}

@inproceedings{zhouleast,
  title={Least-to-Most Prompting Enables Complex Reasoning in Large Language Models},
  author={Zhou, Denny and Sch{\"a}rli, Nathanael and Hou, Le and Wei, Jason and Scales, Nathan and Wang, Xuezhi and Schuurmans, Dale and Cui, Claire and Bousquet, Olivier and Le, Quoc V and others},
  booktitle={The Eleventh International Conference on Learning Representations}
}

@inproceedings{wang2023plan,
  title={Plan-and-Solve Prompting: Improving Zero-Shot Chain-of-Thought Reasoning by Large Language Models},
  author={Wang, Lei and Xu, Wanyu and Lan, Yihuai and Hu, Zhiqiang and Lan, Yunshi and Lee, Roy Ka-Wei and Lim, Ee-Peng},
  booktitle={Proceedings of the 61st Annual Meeting of the Association for Computational Linguistics (Volume 1: Long Papers)},
  pages={2609--2634},
  year={2023}
}

@inproceedings{chengbinding,
  title={Binding Language Models in Symbolic Languages},
  author={Cheng, Zhoujun and Xie, Tianbao and Shi, Peng and Li, Chengzu and Nadkarni, Rahul and Hu, Yushi and Xiong, Caiming and Radev, Dragomir and Ostendorf, Mari and Zettlemoyer, Luke and others},
  booktitle={The Eleventh International Conference on Learning Representations}
}

@article{wei2022chain,
  title={Chain-of-thought prompting elicits reasoning in large language models},
  author={Wei, Jason and Wang, Xuezhi and Schuurmans, Dale and Bosma, Maarten and Xia, Fei and Chi, Ed and Le, Quoc V and Zhou, Denny and others},
  journal={Advances in neural information processing systems},
  volume={35},
  pages={24824--24837},
  year={2022}
}

@inproceedings{sui2024table,
  title={Table meets llm: Can large language models understand structured table data? a benchmark and empirical study},
  author={Sui, Yuan and Zhou, Mengyu and Zhou, Mingjie and Han, Shi and Zhang, Dongmei},
  booktitle={Proceedings of the 17th ACM International Conference on Web Search and Data Mining},
  pages={645--654},
  year={2024}
}

@inproceedings{cotc2024,
  title={Self-Consistency Improves Chain of Thought Reasoning in Language Models},
  author={Wang, Xuezhi and Wei, Jason and Schuurmans, Dale and Le, Quoc V and Chi, Ed H and Narang, Sharan and Chowdhery, Aakanksha and Zhou, Denny},
  booktitle={The Eleventh International Conference on Learning Representations},
  year={2024}
}

@article{reactable2024,
  title={ReAcTable: enhancing ReAct for table question answering},
  author={Zhang, Yunjia and Henkel, Jordan and Floratou, Avrilia and Cahoon, Joyce and Deep, Shaleen and Patel, Jignesh M},
  journal={Proceedings of the VLDB Endowment},
  volume={17},
  number={8},
  pages={1981--1994},
  year={2024},
  publisher={VLDB Endowment}
}

@inproceedings{chainoftable2024,
  title={Chain-of-Table: Evolving Tables in the Reasoning Chain for Table Understanding},
  author={Wang, Zilong and Zhang, Hao and Li, Chun-Liang and Eisenschlos, Julian Martin and Perot, Vincent and Wang, Zifeng and Miculicich, Lesly and Fujii, Yasuhisa and Shang, Jingbo and Lee, Chen-Yu and others},
  booktitle={ICLR},
  year={2024}
}

@inproceedings{mact2025,
  title={Efficient Multi-Agent Collaboration with Tool Use for Online Planning in Complex Table Question Answering},
  author={Zhou, Wei and Mesgar, Mohsen and Friedrich, Annemarie and Adel, Heike},
  booktitle={Findings of the Association for Computational Linguistics: NAACL 2025},
  pages={945--968},
  year={2025}
}

@article{lotus2024,
  title={Semantic Operators and Their Optimization: Enabling LLM-Based Data Processing with Accuracy Guarantees in LOTUS},
  author={Patel, Liana and Jha, Siddharth and Pan, Melissa and Gupta, Harshit and Asawa, Parth and Guestrin, Carlos and Zaharia, Matei},
  journal={Proceedings of the VLDB Endowment},
  volume={18},
  number={11},
  pages={4171--4184},
  year={2025},
  publisher={VLDB Endowment}
}

@inproceedings{tablecritic2025,
    title = "Table-Critic: A Multi-Agent Framework for Collaborative Criticism and Refinement in Table Reasoning",
    author = "Yu, Peiying  and
      Chen, Guoxin  and
      Wang, Jingjing",
    booktitle = "Proceedings of the 63rd Annual Meeting of the Association for Computational Linguistics (Volume 1: Long Papers)",
    month = jul,
    year = "2025",
    address = "Vienna, Austria",
    publisher = "Association for Computational Linguistics",
    pages = "17432--17451",
}

@article{browne2012mcts,
  title={A survey of monte carlo tree search methods},
  author={Browne, Cameron B and Powley, Edward and Whitehouse, Daniel and Lucas, Simon M and Cowling, Peter I and Rohlfshagen, Philipp and Tavener, Stephen and Perez, Diego and Samothrakis, Spyridon and Colton, Simon},
  journal={IEEE Transactions on Computational Intelligence and AI in games},
  volume={4},
  number={1},
  pages={1--43},
  year={2012},
  publisher={IEEE}
}

@article{yao2023tree,
  title={Tree of thoughts: Deliberate problem solving with large language models},
  author={Yao, Shunyu and Yu, Dian and Zhao, Jeffrey and Shafran, Izhak and Griffiths, Tom and Cao, Yuan and Narasimhan, Karthik},
  journal={Advances in neural information processing systems},
  volume={36},
  pages={11809--11822},
  year={2023}
}

@inproceedings{chen2024tree,
  title={When is tree search useful for llm planning? it depends on the discriminator},
  author={Chen, Ziru and White, Michael and Mooney, Ray and Payani, Ali and Su, Yu and Sun, Huan},
  booktitle={Proceedings of the 62nd Annual Meeting of the Association for Computational Linguistics (Volume 1: Long Papers)},
  pages={13659--13678},
  year={2024}
}

@article{chen2025expanding,
  title={Expanding before Inferring: Enhancing Factuality in Large Language Models through Premature Layers Interpolation},
  author={Chen, Dingwei and Liu, Ziqiang and Fang, Feiteng and Leong, Chak Tou and Ni, Shiwen and Argha, Ahmadreza and Alinejad-Rokny, Hamid and Yang, Min and Li, Chengming},
  journal={arXiv preprint arXiv:2506.02973},
  year={2025}
}

@article{ke2024unveiling,
  title={Unveiling factuality and injecting knowledge for LLMs via reinforcement learning and data proportion},
  author={Ke, Wenjun and Shang, Ziyu and Luo, Zhizhao and Wang, Peng and Guo, Yikai and Liu, Qi and Chen, Yuxuan},
  journal={Science China. Information Sciences},
  volume={67},
  number={10},
  pages={209101},
  year={2024},
  publisher={Springer Nature BV}
}

@article{caciularu2024tact,
  title={Tact: Advancing complex aggregative reasoning with information extraction tools},
  author={Caciularu, Avi and Jacovi, Alon and Ben-David, Eyal and Goldshtein, Sasha and Schuster, Tal and Herzig, Jonathan and Elidan, Gal and Globerson, Amir},
  journal={Advances in Neural Information Processing Systems},
  volume={37},
  pages={34775--34799},
  year={2024}
}

@article{psallidas2018smoke,
  title={Smoke: fine-grained lineage at interactive speed},
  author={Psallidas, Fotis and Wu, Eugene},
  journal={Proceedings of the VLDB Endowment},
  volume={11},
  number={6},
  pages={719--732},
  year={2018},
  publisher={VLDB Endowment}
}

@article{haffner2023efficiently,
  title={Efficiently Computing Join Orders with Heuristic Search},
  author={Haffner, Immanuel and Dittrich, Jens},
  journal={Proceedings of the ACM on Management of Data},
  volume={1},
  number={1},
  pages={1--26},
  year={2023},
  publisher={ACM New York, NY, USA}
}

@article{chen2023loger,
  title={Loger: A learned optimizer towards generating efficient and robust query execution plans},
  author={Chen, Tianyi and Gao, Jun and Chen, Hedui and Tu, Yaofeng},
  journal={Proceedings of the VLDB Endowment},
  volume={16},
  number={7},
  pages={1777--1789},
  year={2023},
  publisher={VLDB Endowment}
}

@inproceedings{zhou2024language,
  title={Language agent tree search unifies reasoning, acting, and planning in language models},
  author={Zhou, Andy and Yan, Kai and Shlapentokh-Rothman, Michal and Wang, Haohan and Wang, Yu-Xiong},
  booktitle={Proceedings of the 41st International Conference on Machine Learning},
  pages={62138--62160},
  year={2024}
}

@misc{qwen2.5,
    title = {Qwen2.5: A Party of Foundation Models},
    url = {https://qwenlm.github.io/blog/qwen2.5/},
    author = {Qwen Team},
    month = {September},
    year = {2024}
}

@misc{qwen3technicalreport,
      title={Qwen3 Technical Report}, 
      author={Qwen Team},
      year={2025},
      eprint={2505.09388},
      archivePrefix={arXiv},
      primaryClass={cs.CL},
      url={https://arxiv.org/abs/2505.09388}, 
}

@misc{openai2025gpt41,
  title        = {Introducing GPT-4.1 in the API},
  author       = {{OpenAI}},
  year         = {2025},
  howpublished = {\url{https://openai.com/index/gpt-4-1/}},
  note         = {Accessed: YYYY-MM-DD}
}

@inproceedings{kwon2023efficient,
  title={Efficient Memory Management for Large Language Model Serving with PagedAttention},
  author={Woosuk Kwon and Zhuohan Li and Siyuan Zhuang and Ying Sheng and Lianmin Zheng and Cody Hao Yu and Joseph E. Gonzalez and Hao Zhang and Ion Stoica},
  booktitle={Proceedings of the ACM SIGOPS 29th Symposium on Operating Systems Principles},
  year={2023}
}

@article{paszke2017automatic,
  title={Automatic differentiation in PyTorch},
  author={Paszke, Adam and Gross, Sam and Chintala, Soumith and Chanan, Gregory and Yang, Edward and DeVito, Zachary and Lin, Zeming and Desmaison, Alban and Antiga, Luca and Lerer, Adam},
  year={2017}
}

\end{document}